\documentclass[useAMS,usenatbib]{mn2e}

\usepackage{graphicx,amsmath}
\usepackage{multirow}
\newcommand{\MS}{\ifmmode{\,}\else\thinspace\fi{\rm M}\ifmmode_{\odot}\else$_{\odot}$\fi}
\newcommand{\LS}{\ifmmode{\,}\else\thinspace\fi{\rm L}\ifmmode_{\odot}\else$_{\odot}$\fi}
\newcommand{\Ke}{\ifmmode{\,}\else\thinspace\fi{\rm K}}
\newcommand{\teffs}{\ifmmode T_{\rm eff}^{\rm s}\else$T_{\rm eff}^{\rm s}$\fi}
\newcommand{\setp}{P1}
\newcommand{\setj}{P2}
\newcommand{\setzeo}{P3}
\newcommand{\BLGX}{BLG184.7\thinspace 133264}
\newcommand{\BLGY}{BLG189.6\thinspace 137529}

\title[Period doubling in BL Herculis stars]{Discovery of period doubling in BL Herculis stars of the OGLE survey.
Observations and theoretical models.}
\author[R. Smolec et al.]
{R. Smolec$^{1}$\thanks{E-mail:
radek.smolec@univie.ac.at},
I. Soszy\'nski$^{2}$,
P. Moskalik$^{3}$,
A. Udalski$^{2}$,
M.~K. Szyma\'nski$^{2}$,
\and
M. Kubiak$^{2}$,
G. Pietrzy\'nski$^{2,4}$,
\L{}. Wyrzykowski$^{2,5}$,
K. Ulaczyk$^{2}$,
R. Poleski$^{2}$,
\and
S. Koz\l{}owski$^{2}$ and
P. Pietrukowicz$^{2}$\\
$^{1}$Institute for Astronomy (IfA), University of Vienna, T\"urkenschanzstrasse 17, A-1180 Wien, Austria\\
$^{2}$Warsaw University Observatory, Al. Ujazdowskie 4, 00-478 Warszawa, Poland\\
$^{3}$Copernicus Astronomical Centre, Bartycka 18, 00-716 Warszawa, Poland\\
$^{4}$Universidad de Concepci\'on, Departamento de Astronomia, Casilla 160???C, Concepci\'on, Chile\\
$^{5}$Institute of Astronomy, University of Cambridge, Madingley Road, Cambridge CB3 0HA, UK
}
\begin{document}

\date{Accepted . Received ; in original form }

\pagerange{\pageref{firstpage}--\pageref{lastpage}} \pubyear{2011}

\maketitle

\label{firstpage}

\begin{abstract}

We report the discovery of a period doubling behaviour in a
2.4\thinspace d BL~Herculis-type variable of the Galactic bulge.
Another bulge BL~Her-type star ($P=2.25$\thinspace d) is a strong
period doubling candidate. Both objects have been identified with
the OGLE-III photometry. Possibility of period doubling in this type
of pulsators has been predicted almost twenty years ago by
\cite{bm92}. Our finding is the first observational confirmation of
their theoretical results.

Discovery of the first BL~Herculis star showing the period doubling
effect motivates a new theoretical investigation with the
state-of-the-art convective pulsation codes. We present the results
of initial model survey, specifically aimed at studying the observed
period-doubled BL~Her variable. All of our non-linear models have
$P=2.4$\thinspace d. The computations confirm that the period
doubling effect is caused by the 3:2 resonance between the
fundamental mode and the first overtone, as indicated by
earlier radiative models of Buchler \& Moskalik. Comparison of the
computed and the observed light curves allows to constrain the
parameters of the star, in particular its metallicity, which appears
to be high, $Z\approx 0.01$. The recent evolutionary tracks put also
constraint on the mass of the star, which is close to $M=0.50\MS$.

\end{abstract}


\section{Introduction}\label{sec.intro}

BL~Herculis-type variables are short-period ($1-4$\thinspace d)
subgroup of the Type~II (or Population~II) Cepheids. They are
predominantly found in the old stellar populations, in the globular
clusters and in the Galactic halo. They are believed to be low-mass
stars evolving away from the Zero-Age Horizontal Branch (ZAHB)
towards the Asymptotic Giant Branch (AGB), crossing the instability
strip (IS) at luminosities larger than RR~Lyrae stars \citep[][see
also Section~\ref{sec.evol}]{strom70,g85}. All Type~II
Cepheids, including BL~Her variables, obey a well defined
period-luminosity relation and can be used as distance indicators
\citep[e.g.][and references therein]{maj09a,maj09b}. They are also
key objects for studying the stellar and galactic evolution
\citep[see e.g.,][]{wallerstein}.

Characteristic feature of the BL~Her light curves is a secondary
bump observed in some of the variables. The phenomenon is similar to
the Hertzsprung bump progression observed in the classical,
Population~I Cepheids and is also interpreted as caused by the
resonant interaction between the fundamental mode and the second
overtone \citep{hck82}. Some early hydrodynamic models of BL~Her
variables were computed by e.g. \cite{hck82} and \cite{cs82}.
\cite{ff85} computed several Type~II Cepheid models, including
BL~Her models. One of their models revealed oscillations with
alternating amplitude and period. A large and systematic survey of
the radiative BL~Her models was conducted by \cite{bm92} and
\cite{mb93a}. In most of their model sequences \cite{bm92} found
narrow windows of parameters for which the light and radial velocity
curves exhibited a periodic alternations of deep and shallow minima
-- the period doubling effect. The origin of the period doubling in
hydrodynamic models was traced by \cite{mb90} to the destabilising
role of the half-integer resonances, $(2n+1)\omega_0=2\omega_k$,
between the fundamental mode and the higher order overtones. Close
to the resonance centre period-doubling bifurcation can occur: the
fundamental mode limit cycle becomes unstable, and period-two
solution, with alternating shape of the light and radial velocity
curves can emerge. \cite{bm92} \citep[see also][]{mb93b} showed that
their BL~Her models undergo a period-doubling bifurcation caused by
the 3:2 resonance between the fundamental mode and the first
overtone. \cite{bm92} concluded that period doubling effect, not
observed in any BL Her star at that time, should be eventually
found. More recently, \cite{mdc07} conducted a large survey of
BL~Her-type convective models, however none of their models
exhibited period doubling.

Alternating deep and shallow minima are characteristic feature
of RV~Tau-type variables, which, together with W~Vir-type stars,
form another subgroup of the Type~II Cepheids \citep[for recent
review see e.g.,][]{wallerstein, szabados}. These longer period
variables (periods of $\sim 4 - 20$\thinspace d for W~Vir stars and
above $\sim\! 20$\thinspace d for RV~Tau stars) are more
evolutionary advanced than the BL~Her variables. The W~Vir stars
loop into the instability strip during the helium-shell flashes,
while climbing up along the AGB, whereas the RV~Tau stars enter the
instability strip evolving away from the AGB toward a white dwarf
domain. Period doubling behaviour appears first at $P \simeq
19-20$\thinspace d; this is a borderline between the W~Vir and
RV~Tau classes. However, the two classes overlap in period and the
distinction between them is not clear-cut \citep[e.g.,][]{szabados,
wo08}. The RV~Tau behaviour was reproduced in the radiative
hydrodynamic models of \cite{bk87} and \cite{kb88}. As analysed
later by \cite{mb90}, the period doubling in these models is caused
by the 5:2 resonance between the fundamental mode and the second
overtone.

Period doubling was also discovered in the {\it Kepler} observations
of RR~Lyrae stars exhibiting the Blazhko phenomenon
\citep{kol10,szabo10}. However, to the present day period doubling
was not detected in any BL~Her-type star, despite theoretical
predictions.

In this paper we report the discovery of the first BL~Her
variable clearly showing the period doubling phenomenon as predicted
by the theoretical models. The star was found during the preparation
of the subsequent part of the OGLE-III Catalogue of Variable
Stars (OIII-CVS), devoted to the Type~II Cepheids in the
Galactic bulge (Soszy\'nski et al. 2011, in~prep.). In
Section~\ref{sec.observations} we present and analyse the light
curve of the star and prove that it is indeed a BL~Her-type
pulsator. We also discuss another candidate in which period doubling
is strongly suspected, for which however, definite claims
require more precise photometric observations.

In Section~\ref{sec.models} we present a new set of non-linear
hydrodynamic models. As compared to the \cite{mb93a} models, these
models are convective and include up to date physics. Our efforts
focus on modelling the single BL~Her star with unambiguously
detected period doubling. In majority of our model sequences we find
the period doubling domains and show that they originate from the
resonant interaction of the fundamental mode with the first
overtone, corroborating the earlier result of \cite{bm92}.
Comparison of the model and observed light curves allows to
constrain the parameters of the star (mass, luminosity and
metallicity). In Section~\ref{sec.evol} we confront these parameters
with the predictions of the stellar evolution theory. We discuss our
results and summarise the conclusions in
Section~\ref{sec.discussion}.

\section{Observations and data analysis}\label{sec.observations}

In this study we used Cousins $I$-band photometry of the Galactic
bulge, obtained during the third phase of the Optical Gravitational
Lensing Experiment (OGLE-III). The observations were collected
between 2001 and 2009, with the 1.3\thinspace m Warsaw telescope
located at Las Campanas Observatory, Chile. A limited number of
measurements in Johnson's $V$-band was also obtained to determine
$(V-I)$ colour of the stars. The photometry was reduced with the
Difference Image Analysis method \citep[DIA,][]{al98,woz00}.
Detailed description of the OGLE-III instrumental setup and data
reduction procedures can be found in \cite{ud03} and \cite{ud08}.

During the preparation of the catalogue of Type~II Cepheids in the
Galactic bulge (Soszy\'nski et al. 2011, in prep.), two BL~Her stars
with unusual properties were identified. Both objects display
excessive scatter around the light curve maxima and minima. This
attracted our attention, and prompted thorough investigation of
these two variables.

\subsection{\BLGX\thinspace: first BL~Herculis-type star with period doubling}\label{sec.star1}

\begin{figure}
\centering
\resizebox{\hsize}{!}{\includegraphics{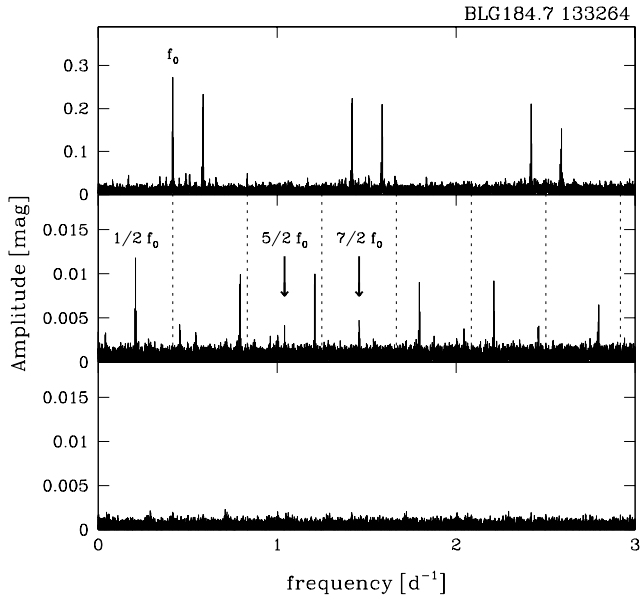}}
\caption{Pre-whitening sequence for \BLGX. Upper panel: power
spectrum of original data. The spectrum is dominated by pulsation
frequency $f_0$ and its daily aliases. Middle panel: power spectrum
after removing $f_0$ and its harmonics. Dashed lines indicate
locations of the pre-whitened frequencies. The highest peak
corresponds to a subharmonic frequency, $1/2\, f_0$, its daily
aliases are very prominent, too. Two other subharmonic frequencies
($5/2\, f_0$ and $7/2\, f_0$) are also visible. Lower panel: power
spectrum after removing $f_0$, its harmonics and its subharmonics.}
\label{fig1a}
\end{figure}

The OGLE-III $I$-band photometry of \BLGX\ (18:02:56.60, $-$30:42:34.4)
consists of 768
measurements, spanning 2830\thinspace d. We analysed these data
using a standard consecutive pre-whitening technique. The star
displays a very pronounced variability with the period of
$P_0=2.399267$\thinspace d ($f_0 = 0.4167939$\thinspace d$^{-1}$). We
fitted this variability with the Fourier series of the form:

\begin{equation}
I(t) = I_0 + \sum_{k} A_{k} \cos (2\pi k f_0 t + \phi_k),
\label{fsum}
\end{equation}

\noindent where the frequency $f_0$ was also optimised. In the next
step, we subtracted the fitted function from the light curve
("pre-whitening"), and then searched the residuals for additional
periodicities. This was done with the standard Fourier power
spectrum. In Fig.~\ref{fig1a} we present the results of this
procedure. The power spectrum of the pre-whitened data (second panel
of the figure) is dominated by a peak at $f_x=0.20841$\thinspace
d$^{-1}$. Within accuracy of the data, this frequency is identical
to $1/2\, f_0$. In other words, $f_x$ is a {\it subharmonic} of the
primary pulsation frequency $f_0$. Subsequent step by step
pre-whitening revealed additional subharmonics, up to $9/2\, f_0$
($5/2\, f_0$ and $7/2\, f_0$ are noticeable already in
Fig.~\ref{fig1a}). No other secondary periodicities were found.

\begin{figure}
\centering
\resizebox{\hsize}{!}{\includegraphics{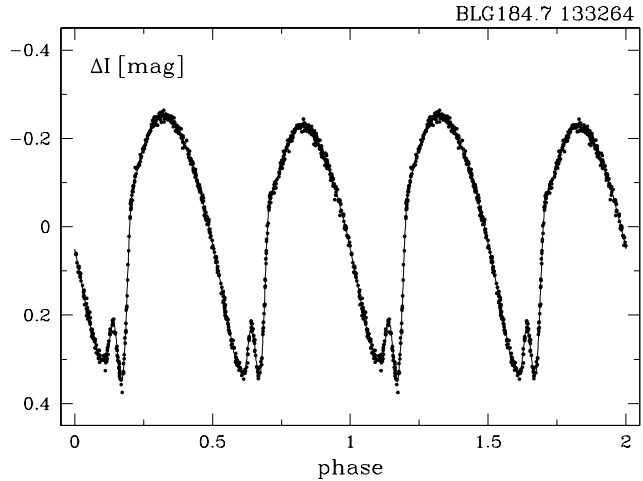}}
\caption{Light curve of \BLGX\ folded with twice the pulsation period, $2P_0$.}
\label{fig2}
\end{figure}

The presence of subharmonic frequencies in the power spectrum, i.e.
frequencies of the form $(n+1/2)f_0$, is a characteristic signature
of a period doubling. In the time domain it means that the light
curve of the star is still {\it strictly periodic}, but it repeats
itself not after one, but after {\it two} pulsation periods. In
other words, the light curve displays regular alternations of higher
and lower maxima (and minima). This is clearly visible in
Fig.~\ref{fig2}, where we plot the light curve of \BLGX\ folded with
twice the pulsation period ($2P_0$). The difference between the two
consecutive maxima is small (0.020\thinspace mag), nevertheless well
visible. Such alternating light curves are well known in longer
period W~Vir and RV~Tau variables \citep[$P_0 >
19-20$\thinspace d, see][]{wo08}, but they were never before
observed in the BL~Her stars. \BLGX\ is the first BL~Her-type
variable, in which period doubling is unambiguously detected.

To fully characterise the light curve of \BLGX, we fitted the data
with the final Fourier series, which included all the detected
subharmonics:

\begin{align}
I(t) = I_0  + &\sum_{k} A_k \cos\big(2\pi k f_0 t + \phi_k\big)+\nonumber \\
+& \sum_{n} B_{n+1/2} \cos\big[2\pi(n+1/2)f_0t + \psi_{n+1/2}\big].
\label{fsums}
\end{align}

\noindent This function is plotted in Fig.~\ref{fig2}, together with the
folded photometric data. The amplitude of the main pulsation
component is $A_{1}=0.2742\pm 0.0005$\thinspace mag. The highest
subharmonic is over twenty times smaller and has an amplitude of
$B_{1/2}=0.0127\pm 0.0005$\thinspace mag. The power spectrum of
residuals of the fit is displayed in the lower panel of Fig.~\ref{fig1a}. The
dispersion of the residuals is $\sigma=0.0086$\thinspace mag.

\begin{figure}
\centering
\resizebox{\hsize}{!}{\includegraphics{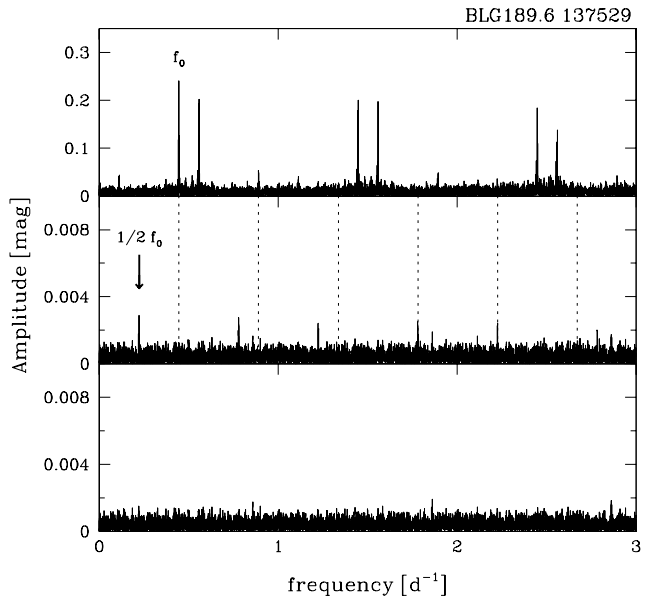}}
\caption{Pre-whitening sequence for \BLGY. Upper panel: power
spectrum of original data. The spectrum is dominated by pulsation
frequency $f_0$ and its daily aliases. Middle panel: power spectrum
after removing $f_0$ and its harmonics. Dashed lines indicate
locations of the pre-whitened frequencies. The highest peak
corresponds to a subharmonic frequency, $1/2\, f_0$, its daily
aliases are also visible. Lower panel: power spectrum after removing
$f_0$, its harmonics and its subharmonics.}
\label{fig3a}
\end{figure}

\begin{figure}
\centering
\resizebox{\hsize}{!}{\includegraphics{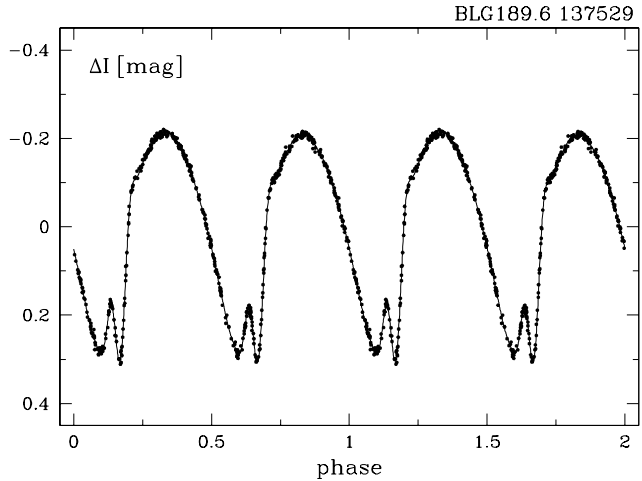}}
\caption{Light curve of \BLGY\ folded with twice the pulsation period, $2P_0$.}
\label{fig4}
\end{figure}

\subsection{\BLGY\thinspace: period doubling candidate}\label{sec.star2}

\BLGY\ (18:00:48.31, $-$29:58:54.4)
is another BL~Her variable in the Galactic bulge, pulsating
with the period of $P_0=2.245688$\thinspace d ($f_0 =
0.4452978$\thinspace d$^{-1}$). During almost 8 years of OGLE-III
observations the pulsation period of the star was slowly changing.
Therefore, we decided to use in the analysis only a subset of data
spanning 1540\thinspace d, from HJD\, =\, 2453416 to 2454956 (543
measurements in total). The data was analysed in the same way as in
the previous case.

The pre-whitening sequence of \BLGY\ is displayed in
Fig.~\ref{fig3a}. Only one secondary peak was detected in this star.
Its frequency, $f_x=0.22257$\thinspace d$^{-1}$, corresponds $1/2\,
f_0$. Thus, \BLGY\ shows a signature of a period doubling. This time
however, the detected subharmonic is extremely weak, with the
amplitude of only $B_{1/2}=0.0029\pm 0.0004$\thinspace mag. The
light curve of \BLGY, folded with twice the pulsation period is
displayed in Fig.~\ref{fig4}. As we might have expected, the
alternations of maxima and of minima are barely visible.

\begin{figure}
\centering
\resizebox{\hsize}{!}{\includegraphics{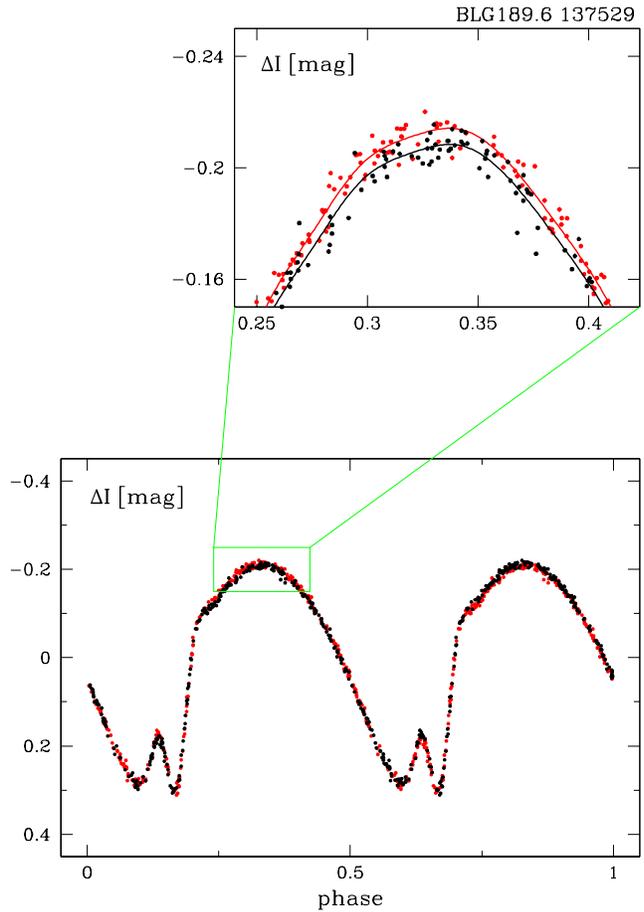}}
\caption{Light curve of \BLGY\ folded with $2P_0$, plotted twice with
different colours. The red and black curves are shifted by $P_0$. The
inset shows the difference between even and odd pulsation cycles.}
\label{fig5}
\end{figure}

To make them visible better, we display the same data again in
Fig.~\ref{fig5}, but this time we do it differently. We plot the
folded data twice, with different colours. The two curves are
shifted by one pulsation period (i.e. by $P_0$). In this way, even
and odd pulsation cycles are plotted on top of each other. The inset
provides a closer look at the maxima of the folded curves. The red
points are systematically higher than the black ones, but the
difference is very small. The maxima of the fit lines differ by only
0.0058\thinspace mag, which is less than the scatter of the data
around the fit (0.0064\thinspace mag).

\begin{figure}
\centering
\resizebox{\hsize}{!}{\includegraphics{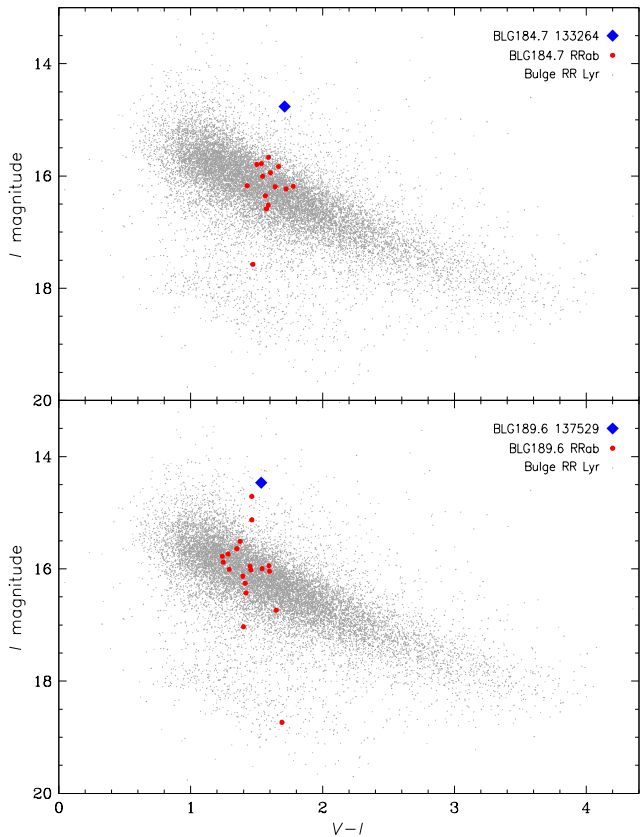}}
\caption{Colour-magnitude diagram for \BLGX\ and \BLGY\ (plotted
with diamonds). Background grey dots represent RR~Lyr stars observed
toward the Galactic bulge. Filled circles mark RRab stars observed
in the same fields as \BLGX\ and \BLGY.}
\label{fig6}
\end{figure}

\begin{figure}
\centering
\resizebox{\hsize}{!}{\includegraphics{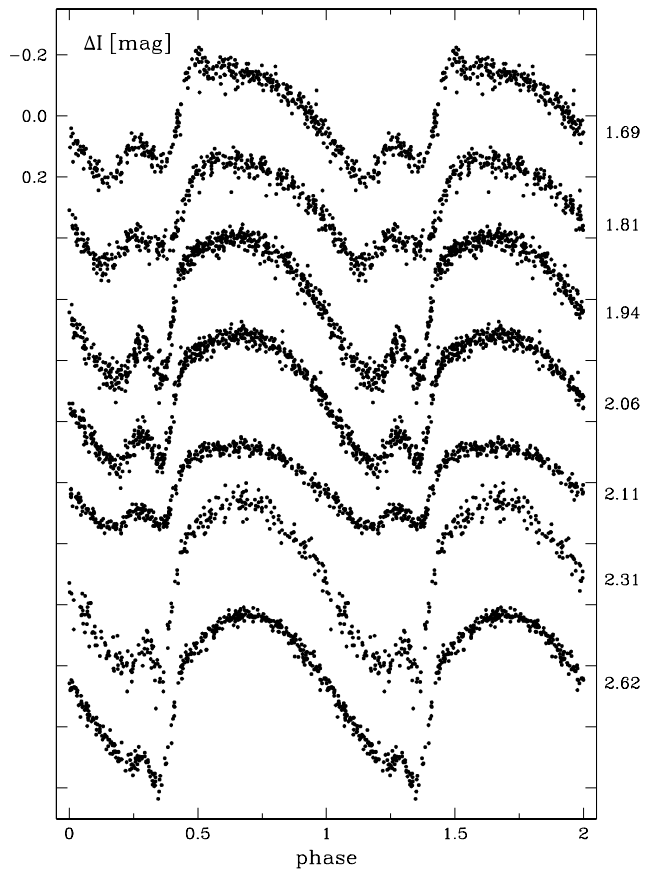}}
\caption{Light curves of selected BL~Her stars of the LMC
\citep{sosz08}. Numbers at the right side of the plot show periods
(in days) of the displayed variables.}
\label{fig7}
\end{figure}

Both Fig.~\ref{fig3a} and Fig.~\ref{fig5} show that the period
doubling in \BLGY, is present, but is much less significant than in
the first star. We consider this BL~Her variable to be a strong
period doubling candidate, which however needs further confirmation
with additional data. Such data are currently being collected
(for both stars of this paper) within the fourth phase of the OGLE
project.

\subsection{Are \BLGX\ and \BLGY\ genuine BL~Her-type variables ?}\label{sec.BLHer}

In Fig.~\ref{fig6} we put \BLGX\ and \BLGY\ on the colour-magnitude
diagram, together with almost 17000 RR~Lyr variables observed toward
the Galactic bulge \citep{sosz11}. The bulge RR~Lyr stars are spread
over a wide range of $(V-I)$ colours. This is caused by large
field-to-filed variations of interstellar reddening. We plot
with the filled circles the RRab stars located in the same fields
as \BLGX\ and \BLGY. Both period-doubled variables discussed in the
current paper have the same colours as the RRab stars in their
respective fields (i.e. RRab stars with the same reddening).
This implies that \BLGX\ and \BLGY\ are located inside the
instability strip. Their luminosities, which are $1.65\pm
0.55$\thinspace mag brighter than the Horizontal Branch, place them
in the domain of the BL~Her-type pulsators.

\begin{table}
\caption{Light curves of \BLGX\ and \BLGY.}
\label{tab.fourier}
\centering
\begin{tabular}{lcc}
\hline
                           & \BLGX    & \BLGY \\
\hline
$P_0$\thinspace [d]      & 2.399267 & 2.245688 \\
$A_1$\thinspace [mag]      & 0.2742   & 0.2438   \\
$B_{1/2}$\thinspace [mag]  & 0.0127   & 0.0029   \\
\smallskip
$\sigma$\thinspace [mag]   & 0.0086   & 0.0064   \\
$R_{21}$                   & 0.216    & 0.190    \\
$R_{31}$                   & 0.076    & 0.071    \\
$\phi_{21}$                & 5.461    & 5.590    \\
$\phi_{31}$                & 3.994    & 3.908    \\
\hline
\end{tabular}
\end{table}

\begin{figure*}
\centering
\resizebox{\hsize}{!}{\includegraphics{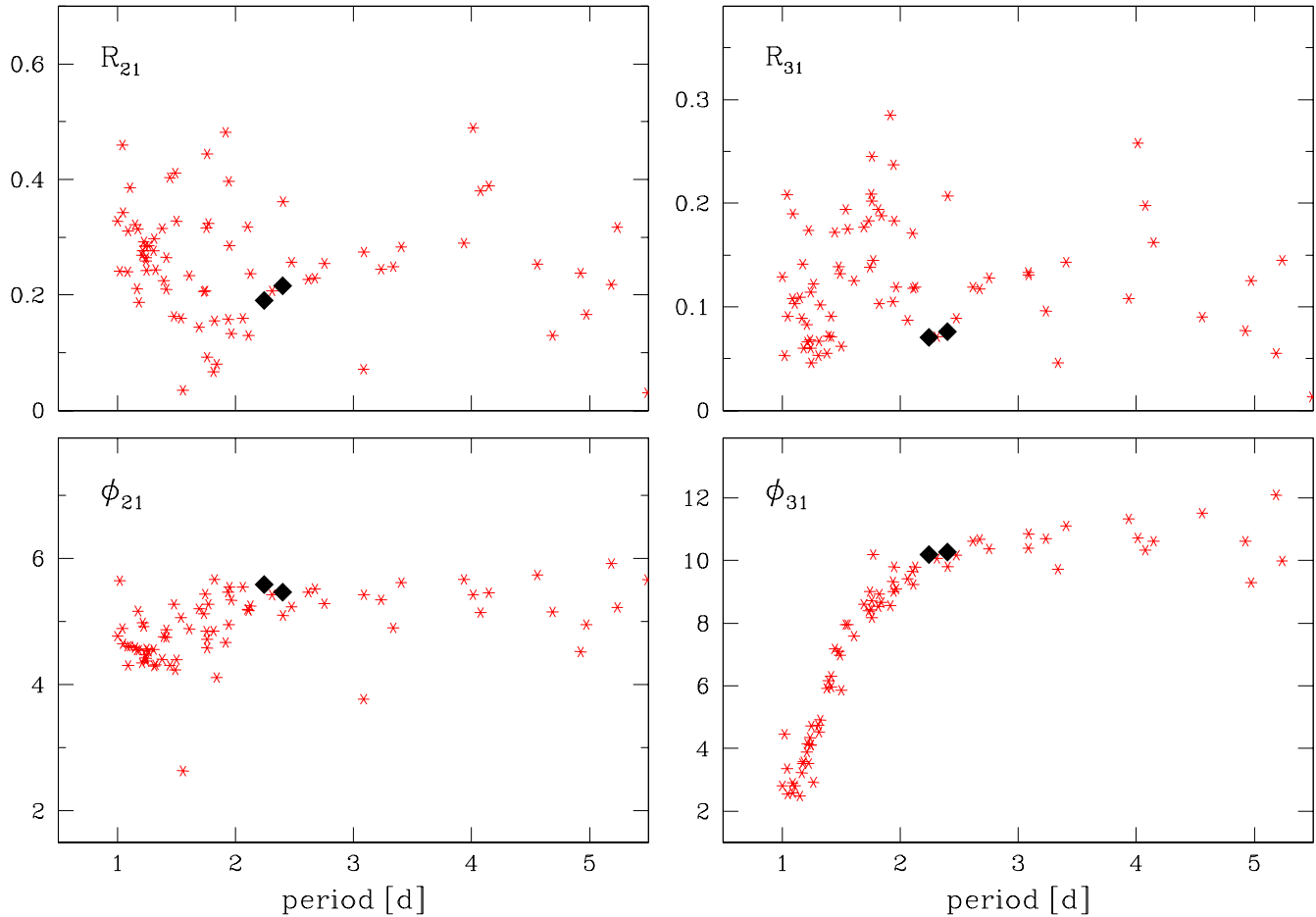}}
\caption{$I$-band Fourier parameter progressions for the Type~II
Cepheids. Asterisks represent Type~II Cepheids of the LMC
\citep[][]{sosz08}. \BLGX\ and \BLGY\ are plotted with diamonds.
Their $\phi_{31}$ values are shifted up by $2\pi$.}
\label{fig8}
\end{figure*}

The light curves of both stars are also quite typical for the BL~Her
variables. In Fig.~\ref{fig7} we display a selection of BL~Her stars
discovered in the Large Magellanic Cloud \citep{sosz08}. All these
variables were identified on the basis of their position in the
$P-L$ diagram, which established their BL~Her nature beyond doubt.
Fig.~\ref{fig7} shows that there is nothing unusual about the light
curves of \BLGX\ and \BLGY. In fact, they are very similar to the
light curves of BL~Her stars observed in the LMC. This statement can
be put on a more quantitative basis with the help of the Fourier
decomposition parameters $R_{k1}=A_k/A_1$ and
$\phi_{k1}=\phi_k-k\phi_1$, where the $A_k$ and $\phi_k$ are defined
by Eq.~\ref{fsum} or Eq.~\ref{fsums}. The values of $R_{k1}$ and
$\phi_{k1}$ for the light curves of \BLGX\ and \BLGY\ are given in
Tab.~\ref{tab.fourier}. In Fig.~\ref{fig8} these values are compared
with the Fourier parameters of the Type~II Cepheids of the LMC.
Both period-doubled variables are firmly located on the Fourier
parameter progressions defined by the LMC pulsators. This is
particularly striking in the plot of the $\phi_{31}$ vs. period,
where Type~II Cepheids form a very tight, well defined sequence.
Based on Figs~\ref{fig6}, \ref{fig7} and \ref{fig8}, we conclude
that \BLGX\ and \BLGY\ are perfectly normal BL~Her-type variables
and, except of the period doubling, they do not differ
in any way from other pulsators of this class.

\section{Hydrodynamic models}\label{sec.models}

The discovery of the first BL~Herculis star showing the period
doubling motivates a new survey of convective hydrodynamic models
including up to date physics. In this Section we report the results
of such initial model survey in which our goals are the following:
({\it i}) to check whether the period doubling is possible in the
convective BL~Her models with the fundamental mode period equal to
2.4\thinspace d; ({\it ii}) to confirm the crucial role of the 3:2
resonance, $3\omega_0=2\omega_1$, evident from the earlier radiative
models of \cite{bm92}; ({\it iii}) to model the light curve of the
only firm BL~Her star showing the period doubling and ({\it iv}) to
constrain its physical parameters (mass, luminosity, metallicity)
based on the comparison of the computed models with observations.

In Section~\ref{sec.numericalmethods} we briefly present the
convective hydrocodes used in this study and provide the details on
the adopted model parameters. Results of the model survey are
presented in the following sections, first, the results of extensive
linear model survey (Section~\ref{sec.linear}), and next, the
results of non-linear model computations focused on reproducing the
light curve of \BLGX\ (Section~\ref{sec.nonlinear}).

\subsection{Numerical methods}\label{sec.numericalmethods}

In all model computations we use the convective pulsation hydrocodes described by \cite{sm08}. The codes are Lagrangian. Radiative transfer is treated in the diffusion approximation. For the convective energy transfer we use the \cite{ku86} time-dependent model adopted for the use in the pulsation hydrocodes. Kuhfu\ss{} model is a simple, one dimensional formulation, with one additional equation describing the generation of the turbulent energy. All equations and details are given in \cite{sm08}. The model contains eight order-of-unity scaling parameters. These are the mixing-length parameter, $\alpha$, and parameters scaling the turbulent fluxes and the terms that drive/damp the turbulent energy, $\alpha_{\rm p}$ (turbulent pressure), $\alpha_{\rm m}$ (eddy-viscous dissipation), $\alpha_{\rm c}$ (convective heat flux), $\alpha_{\rm t}$ (kinetic turbulent energy flux), $\alpha_{\rm s}$ (buoyant driving), $\alpha_{\rm d}$ (turbulent dissipation) and $\gamma_{\rm r}$ (radiative cooling). Theory provides no guidance for their values however, some standard values are in use. They result from comparison of a static, time-independent version of the model with the standard mixing-length theory \citep[see][]{wf98,sm08}. In practice, values of the parameters are adjusted to get the models that satisfy as many observational constraints as possible. Parameters used in this study are summarised in Table~\ref{tab.convpar}. In this initial study (see Section~\ref{sec.discussion}) we neglect the effects of turbulent pressure and turbulent overshooting. Set~\setp\ is our basic choice.

\begin{table}
\caption{Convective parameters considered in this study. Parameters $\alpha_{\rm s}$, $\alpha_{\rm c}$, $\alpha_{\rm d}$, $\alpha_{\rm p}$ and $\gamma_{\rm r}$ are given in the units of standard values \citep[$\alpha_{\rm s}=\alpha_{\rm c}=1/2\sqrt{2/3}$, $\alpha_{\rm d}=8/3\sqrt{2/3}$, $\alpha_{\rm p}=2/3$ and $\gamma_{\rm r}=2\sqrt{3}$; see][for details]{sm08}.}
\label{tab.convpar}
\centering
\begin{tabular}{ccccccccc}
\hline
set & $\alpha$ & $\alpha_{\rm m}$ & $\alpha_{\rm s}$ & $\alpha_{\rm c}$ & $\alpha_{\rm d}$ & $\alpha_{\rm p}$ & $\alpha_{\rm t}$ & $\gamma_{\rm r}$ \\
\hline
\setp   & 1.5 & 0.25 & 1.0 & 1.0 & 1.0 & 0.0 & 0.0 & 0.0\\
\setj   & 1.8 & 0.25 & 1.0 & 1.0 & 1.0 & 0.0 & 0.0 & 0.0\\
\setzeo & 1.5 & 0.85 & 1.0 & 1.0 & 1.0 & 0.0 & 0.0 & 1.0\\
\hline
\end{tabular}
\end{table}

Static envelope models were constructed using 180 mass shells, extending down to $2\!\times\! 10^6\,{\rm K}$, with the fixed temperature ($T_{\rm a}\!=\!15\,000\,{\rm K}$) in the anchor zone located 70 zones below the surface. Such mesh structure, with large number of zones between the hydrogen ionisation region and the surface, allows for better resolution of the convective zone in the non-linear computations, which becomes very thin during the expansion phases, sweeping inward through the Lagrangian zones of the model.

The static model is perturbed (``kicked'') with the scaled linear velocity eigenvector and time evolution is followed. By default, non-linear model integration is conducted for $1\,000$ pulsation cycles. With typical linear growth rates characterising BL~Her stars, on order of $0.1\,{\rm d}^{-1}$, the limit cycle pulsation is reached only after several tenths of pulsation periods.

Non-linear models of BL~Her stars, are numerically much more demanding than the models of RR~Lyrae stars. Having roughly the same mass, BL~Her stars are 2--3 times more luminous. Extended envelope is strongly non-adiabatic and pulsations are rather violent. Consequently artificial viscosity, for which we use the von Neumann and Richtmyer prescription \citep{av57}, turns on in the models. We adopt the following values for the artificial viscosity parameters \citep[see][]{stel75}, $C_{\rm q}=4.0$ and a large cut-off parameter $\alpha_{\rm cut}=0.1$. Large value of the cut-off parameter \citep[$0.1$ as compared to $0.01$ used by default in the radiative computations, see e.g.][]{bm92} assures that the artificial viscosity plays a role only in the regions of the strongest velocity gradients, located in a few outermost layers of our models, and consequently does not affect the dynamics of pulsation\footnote{For the models with the highest $L/M$ ratio artificial viscosity can contribute to the pulsation damping also  in the hydrogen-helium ionisation region, but the contribution is small as compared to the damping caused by the turbulent viscosity.}.

In order to compare the models with the OGLE $I$-band light curve of  \BLGX, the bolometric light curve is transformed to the $I$-band by computing the bolometric correction and $(V-I)$ colour at each pulsation phase. To this aim static \cite{kurucz} atmosphere models\footnote{http://kurucz.harvard.edu/} are used.

\subsection{Linear model survey}\label{sec.linear}

In our model survey we consider four values of model masses, $M=0.50\MS$, $0.55\MS$, $0.60\MS$ and $0.65\MS$. The adopted luminosities are in large range of $90-280\LS$. In all our models the hydrogen abundance is fixed, $X=0.76$, and three different heavy element abundances are considered, $Z=0.0001$, $0.001$ and $0.01$. The OP opacities \citep{sea05} generated for the solar mixture of \citet{a04} are used. At the low temperatures they are supplemented with the \citet{af94} opacity data, which is available for the \cite{gn93} mixture\footnote{The OP and Alexander \& Ferguson opacities are stitched together at $\log T=3.95$, that is in the outer model layers, above partial ionization regions driving the pulsation. Consequently the mismatch in the abundance mixtures does not affect the model behaviour.}. Parameters of the models are not bounded by the evolutionary constraints. Later, in Section~\ref{sec.evol}, we compare our models with the recent horizontal branch evolutionary tracks.

\begin{figure}
\centering
\resizebox{\hsize}{!}{\includegraphics{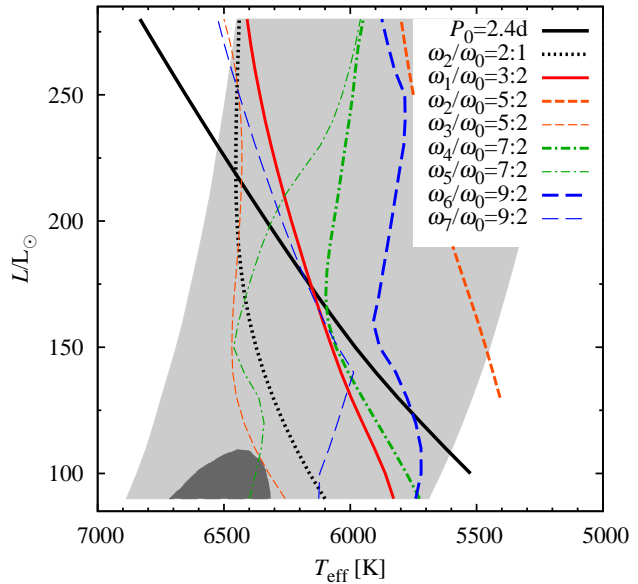}}
\caption{Results of the linear model survey adopting the convective parameters of set \setp\ (Table~\ref{tab.convpar}) and for models with $M=0.55\MS$ and $Z=0.001$. The shaded areas show the instability strips for the fundamental mode (light grey) and for the first overtone (dark grey). The thick solid line is the line of constant fundamental mode period, $P_0=2.4$\thinspace d, and other lines show the loci of various resonances, as indicated in the key. Presented results are based on the interpolation in the extensive grid of linear models with 10\LS\ step in the luminosity and 50\thinspace K step in the effective temperature.}
\label{fig.diags}
\end{figure}

\begin{figure*}
\centering
\resizebox{\hsize}{!}{\includegraphics{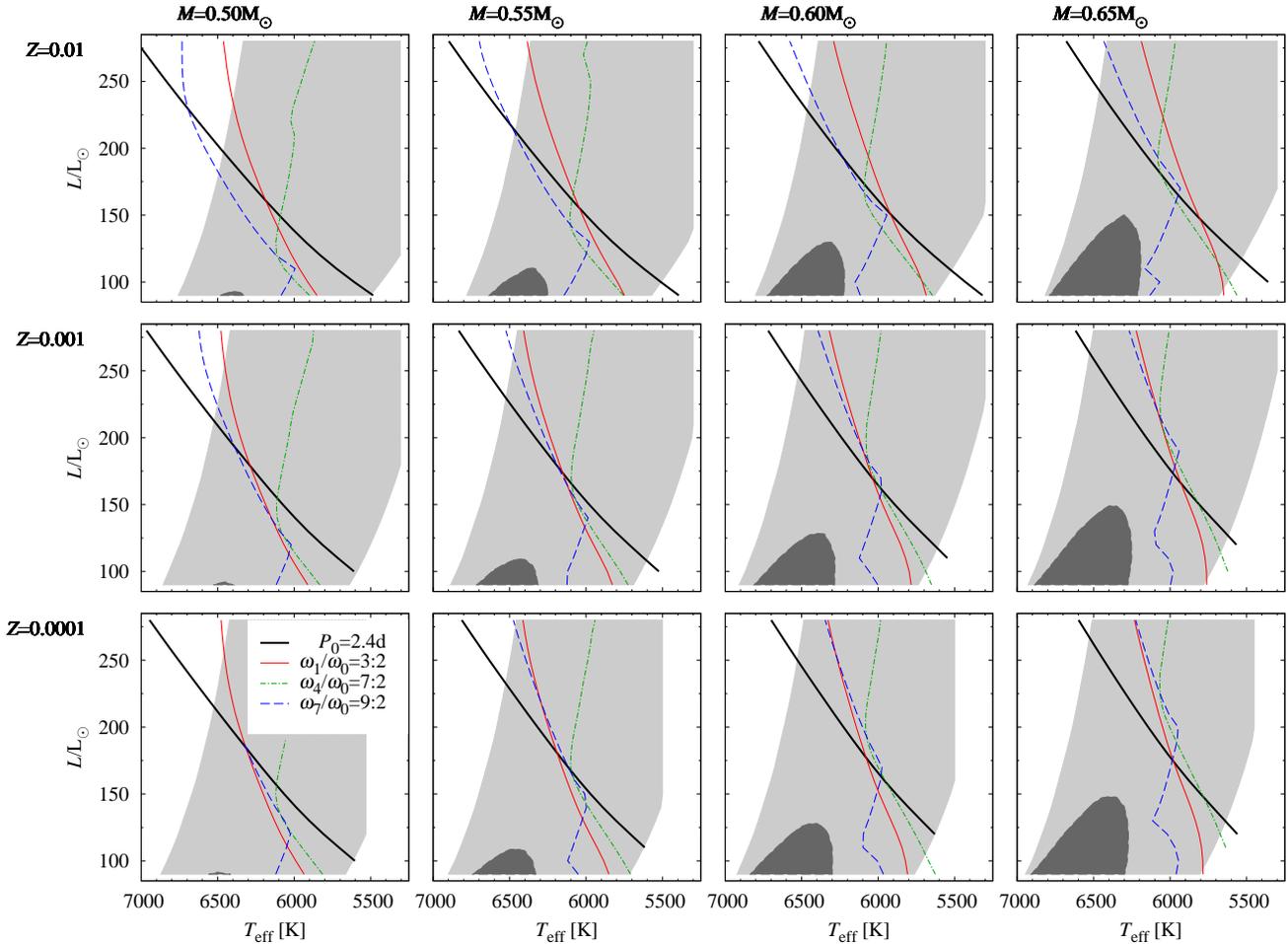}}
\caption{The same as Fig.~\ref{fig.diags} but for a grid of models with different masses ($0.50\MS$, $0.55\MS$, $0.60\MS$, and $0.65\MS$, in consecutive columns), and metallicities ($0.01$, $0.001$, and $0.0001$, in consecutive rows). For clarity, the loci of only three half-integer resonances are plotted, as indicated in the key in the bottom, left-most panel.}
\label{fig.diag5}
\end{figure*}

Results of the linear model survey are shown in the HR diagrams presented in Figs~\ref{fig.diags} and \ref{fig.diag5}. Fig.~\ref{fig.diags} which shows the results for $M=0.55\MS$ and $Z=0.001$ is more detailed than Fig.~\ref{fig.diag5} in which we show the results for the full grid of considered masses and metallicities. In these models convective parameters of set \setp\ were used. In both figures the instability strips are shown as shaded areas, light grey for the fundamental mode, and dark grey for the first overtone. Typical width of the instability strip is on order of $1\ 000$\thinspace K. First overtone can be unstable only for the lowest luminosities. Thick solid line running across the instability strip is the line of constant fundamental mode period equal to 2.4\thinspace d (nearly equal to the period of \BLGX). Thick dotted line (plotted only in Fig.~\ref{fig.diags}) shows the loci of the 2:1 resonance $2\omega_0=\omega_2$ which is crucial in shaping the bump-progression observed in the light curves of BL~Her stars (see Section~\ref{sec.intro}). Other lines show the loci of various half-integer resonances, which are most important in the context of period doubling phenomenon.

 In our study we considered all possible half-integer resonances between the fundamental mode and the eight lowest order overtones. Loci of all these resonances are plotted in Fig.~\ref{fig.diags}. Six resonances cross the $P_0=2.4$\thinspace d line at different luminosities and in principle each of them can be responsible for the period doubling observed in \BLGX. The most likely candidate is the 3:2 resonance with the first overtone, $3\omega_0=2\omega_1$. This resonance was proofed to be responsible for the period doubling in the radiative BL~Her models of \cite{bm92}. Other resonances however, cannot be excluded {\it a priori}. Very promising are the high-order resonances with the higher-order overtones, with the sixth and with the seventh order overtones. These modes are special, as they are trapped in the outer envelope, between the surface and the hydrogen ionisation front. Origin and properties of these modes were studied in detail by \cite{byk97} and \cite{bk01}. Such modes are characterised by a large growth rates as compared to their not-trapped neighbours. In their study of the half-integer resonances \cite{mb90} showed, that period doubling behaviour is possible if the resonant mode is not damped too strongly (but it does not have to be unstable). Consequently, resonances involving the trapped modes are also the likely candidates to cause the period doubling. In our models the growth rates of the sixth and seventh overtones are close to zero, but the modes remain stable. In addition the modes undergo the avoided crossing, which may be noticed from the run of resonant lines in Fig.~\ref{fig.diags} (long-dashed lines). We also note that \cite{kms11} showed that the period doubling discovered recently in several RR~Lyrae stars showing the Blazhko effect \citep{szabo10} is caused by the 9:2 resonance with the ninth order overtone, which is a trapped envelope mode in their models. Also our model computations for the Blazhko RR~Lyrae stars \citep{smk11} confirm the crucial role of this resonance.

Fig.~\ref{fig.diag5} illustrates how the location of the resonance loci vary with mass and metallicity of the models. For clarity only the three half-integer resonances are plotted in the Figure, the 3:2 resonance with the first overtone, $3\omega_0=2\omega_1$, the 7:2 resonance with the fourth overtone, $7\omega_0=2\omega_4$, and the 9:2 resonance with the seventh overtone (trapped mode), $9\omega_0=2\omega_7$. These three resonances cross the $P_0=2.4$\thinspace d line well within the instability strip for almost all considered combinations of $M$ and $Z$. Other resonances often fall beyond or close to the edge of the instability strip. It is visible that changes in the location of resonant lines are rather small, but systematics is clear. Considering e.g. the 3:2 resonance and its crossing with $P_0=2.4$\thinspace d line, one observe that larger the mass (at fixed $Z$) lower the luminosity and lower the effective temperature at the crossing. The same tendency is observed with the increasing metallicity (at fixed mass). The largest possible luminosity at the crossing is $L\approx 185\LS$ at $T_{\rm eff}\approx 6316$\thinspace K  ($M=0.50\MS$, $Z=0.0001$), while the lowest possible luminosity is $L\approx 147\LS$ at $T_{\rm eff}\approx 5804$\thinspace K ($M=0.65\MS$, $Z=0.01$). We note that model periods depend only weakly on the convective parameters of the models. Sensitivity is larger for the period ratios and thus, also for the location of the resonant lines. The quantitative picture arising from Figs~\ref{fig.diags} and \ref{fig.diag5} however, is the same for other sets of convective parameters, in particular for sets \setj~and \setzeo~considered in this study.

Linear computations show, that at least several half-integer resonances can play a role in destabilising the fundamental mode and giving rise to the period doubling behaviour in a models with fundamental mode period of $2.4$\thinspace d. However proximity to the resonance centre and weak damping of the resonant overtone are not sufficient conditions. The non-linear effects which determine the mode interaction are crucial here and only full amplitude non-linear computations can reveal whether and which resonance can cause the period doubling. \\

\subsection{Non-linear model survey}\label{sec.nonlinear}

In this Section we present the results of non-linear model survey aimed to model the light curve of \BLGX, in particular, to reproduce the period doubling behaviour and to investigate its causes. All model sequences run across the fundamental mode instability strip, along the lines of constant period equal to $2.4$\thinspace d (Figs~\ref{fig.diags} and \ref{fig.diag5}). Thirty six model sequences were computed, for all considered values of model masses ($0.50\MS$, $0.55\MS$, $0.60\MS$ and $0.65\MS$), model metallicities ($0.01$, $0.001$ and $0.0001$) and for the three sets of convective parameters (\setp, \setj\ and \setzeo, Table~\ref{tab.convpar}). Individual models in a sequence are specified by the static luminosity, which is changed in the $2\LS$ steps. Effective temperature is adjusted to match the $P_0=2.4$\thinspace d (for considered mass and metallicity). We note that full amplitude non-linear periods may slightly differ from the linear value and in our models are usually longer by up to 3\thinspace per cent.

Bolometric light curves are transformed to the $I$-band (see
Section~\ref{sec.numericalmethods}) and next, are fitted with the
Fourier series (equation \ref{fsums}). We also compute
the Fourier decomposition parameters of low order, the amplitude
ratios, $R_{k1}$, and phase differences, $\phi_{k1}$ (see Section~\ref{sec.BLHer}).

\begin{figure}
\centering
\resizebox{\hsize}{!}{\includegraphics{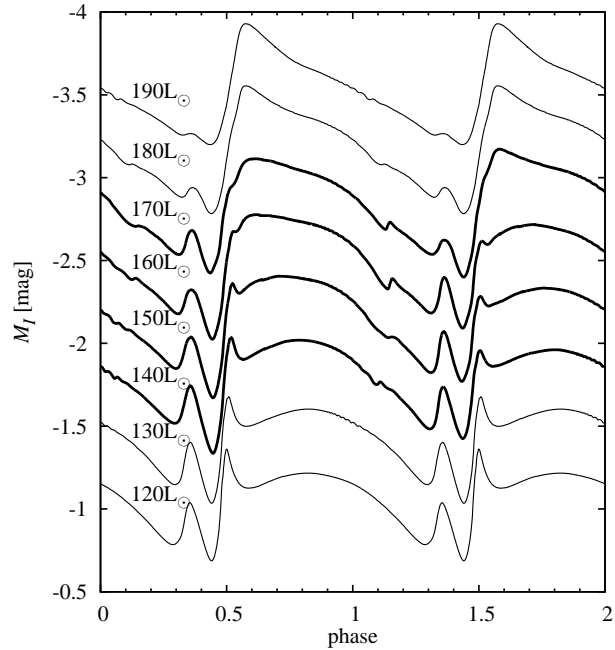}}
\caption{A sample of $I$-band light curves from the model sequence with $M=0.50\MS$ and $Z=0.01$ (set \setp). For clarity, the consecutive light curves are shifted arbitrarily by $0.3$\thinspace mag. Each light curve is marked with luminosity  value on the left-hand side. Light curves showing the period doubling phenomenon ($140-170\LS$) are plotted with thick lines.}
\label{fig.curves}
\end{figure}

Exemplary $I$-band light curves from model sequence with $M=0.50\MS$
and $Z=0.01$ (set \setp) are shown in Fig.~\ref{fig.curves}.
Pulsation period is roughly the same for all these models,
$P_0\approx 2.4$\thinspace d. Systematic change of the light-curve
shapes is clearly visible. Model with the largest luminosity
($190\LS$) has a regular triangular shape almost void of secondary
features. As luminosity decreases the secondary features such as
bumps, become more and more pronounced. Of great interest is the
bump close to the minimum brightness, which is also visible in the
light curve of \BLGX\ (Fig.~\ref{fig2}), as well as in some other
BL~Her-type variables (e.g. Fig.~\ref{fig7}). Its amplitude
increases with the decreasing luminosity. Another bump is visible
close to the maximum brightness and is most pronounced for the
lowest luminosity models. One may also note a series of wiggles of
very low amplitude on the descending branch. It is a numerical
effect caused by the lack of resolution in the convective zone which
becomes very thin during the expansion phases (close to maximum
radius, see also Section~\ref{sec.numericalmethods}). Period
doubling is clearly visible in several models with luminosities in a
range $140-170\LS$.

For other model sequences we observe qualitatively the same picture.
In nearly all our model sequences period doubling behaviour is
evident for several models. In Section~\ref{sec.pd} we focus on the
origin of this phenomenon. Later in Section~\ref{sec.lcc} we focus
on the light-curve shapes and search for the best-fitting model for
\BLGX.

\subsubsection{Origin of the period doubling}\label{sec.pd}

Period doubling is caused by the destabilisation of the fundamental
mode by the resonant overtone. The stability of the limit cycle
pulsation can be studied using the relaxation technique
\citep{stel74}, which shows whether the non-linear limit cycle is
stable or not with respect to the possible perturbation in other
(resonant) mode. Such analysis was done by \cite{bm92} for their
radiative BL~Her models. The period doubling domain they found
coincide with the 3:2 resonance centre. Using the relaxation
technique they showed that for these models the fundamental mode
limit cycle is unstable with respect to the first overtone, pointing
to the crucial role played by the resonance.

Unfortunately, the relaxation scheme is not implemented in our convective hydrocodes. Consequently, we are left to judge on the role played by particular half-integer resonance in the computed period doubling behaviour, based on the distance of the model to the resonance centre (mismatch parameter, $\Delta$), which is computed using the linear period ratios. One have to remember that when non-linear results are correlated with the linear properties of the models, the models have to be computed in a consistent way, i.e. using the same numerical mesh and the same input physics \citep[see e.g.][]{bmk90}. This is the case of our models.

\begin{figure}
\centering
\resizebox{\hsize}{!}{\includegraphics{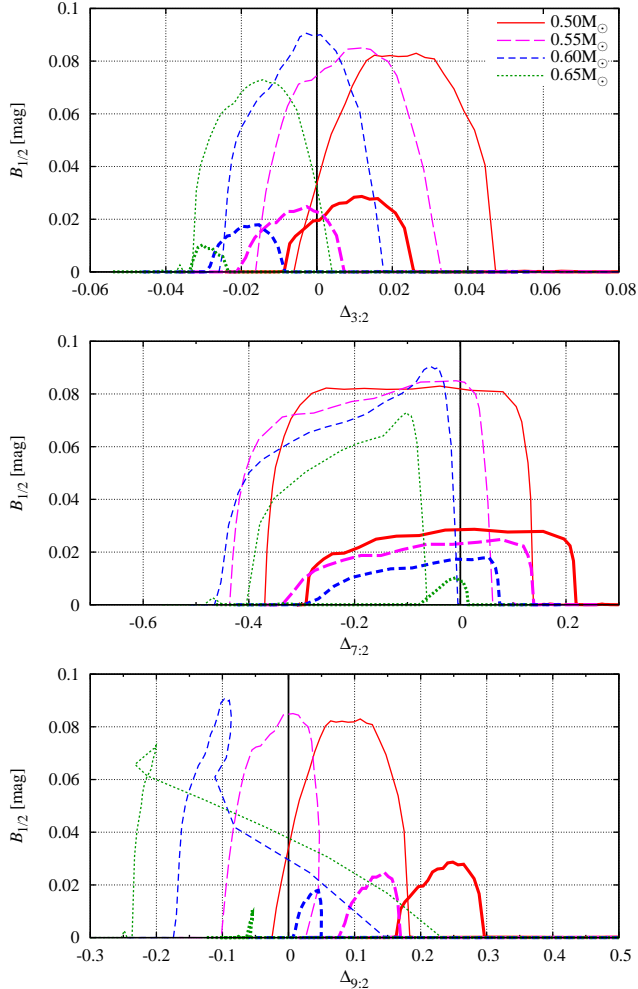}}
\caption{Amplitude of the sub-harmonic frequency component, $B_{1/2}$, plotted versus the mismatch parameters for: (top panel) 3:2 resonance with the first overtone, $\Delta_{3:2}=\omega_1/\omega_0-1.5$, (middle panel) 7:2 resonance with the fourth overtone, $\Delta_{7:2}=\omega_4/\omega_0-3.5$ and, (bottom panel) 9:2 resonance with the seventh overtone, $\Delta_{9:2}=\omega_7/\omega_0-4.5$. Models with different masses are plotted with different line styles as indicated in the key. Metallicity is indicated with the line thickness; the thick lines are for the largest metallicity, $Z=0.01$ and the thin lines are for the lowest metallicity, $Z=0.0001$. For clarity, model sequences with $Z=0.001$ are not shown in the Figure. In all models convective parameters of set \setp\ (Table~\ref{tab.convpar}) were adopted. }
\label{fig.REZ}
\end{figure}

The existence and strength of the period doubling behaviour in our
models is characterised by the amplitude of the sub-harmonic
frequency, $B_{1/2}$ (eq.~\ref{fsums}). In Fig.~\ref{fig.REZ}
$B_{1/2}$ is plotted versus the mismatch parameter for the three
half-integer resonances, $3\omega_0=2\omega_1$
($\Delta_{3:2}=\omega_1/\omega_0-1.5$) in the top panel,
$7\omega_0=2\omega_4$ ($\Delta_{7:2}=\omega_4/\omega_0-3.5$) in the
middle panel, and $9\omega_0=2\omega_7$
($\Delta_{9:2}=\omega_7/\omega_0-4.5$) in the bottom panel. Each
line in this figure corresponds to one model sequence with given
mass and metallicity. All model sequences displayed in
Fig.~\ref{fig.REZ} adopt convective parameters of set \setp. For all
model sequences we see a single period doubling domain with typical
amplitudes of few hundreds of magnitude.

In the top panel of Fig.~\ref{fig.REZ} we clearly see that the
period doubling domains always coincide with the
$3\omega_0=2\omega_1$ resonance centre.  In most cases, the
resonance centre is located within the period doubling domain. For
all models showing the period doubling, the mismatch parameter is
always smaller than $0.05$ (and is usually much smaller). We note
that period doubling domain can be significantly shifted with
respect to the resonance centre, as it is e.g. for $M=0.60\MS$ and
$M=0.65\MS$ ($Z=0.01$) domains which lie outside the resonance
centre. It does not disproof the role played by the resonance
however, but can be caused by the non-linearity and non-adiabaticity
of the pulsations \citep[see e.g.][]{ bm92,rs09a}.

The period doubling domains are also close to the
$7\omega_0=2\omega_4$ resonance centre (middle panel of
Fig.~\ref{fig.REZ}). In most cases the resonance centre falls
somewhere within the period doubling domain. But contrary to the
$3\omega_0=2\omega_1$ resonance, the range of mismatch parameters is
very large, $-0.5<\Delta_{7:2}<0.2$. It is highly unlikely that the
period doubling is caused by the $7\omega_0=2\omega_4$ resonance in
the model located as far from the resonance centre as e.g.
$|\Delta_{7:2}|>0.2$ \citep{mb90}.

For the $9\omega_0=2\omega_7$ resonance (bottom panel of
Fig.~\ref{fig.REZ}) conclusions are the same. Most of the period
doubling models are located far from the resonance centre. Even the
entire period doubling domain can be located as far as
$\Delta_{9:2}>0.15$ (for model sequence with $M=0.50\MS$ and
$Z=0.01$ -- thick solid line in the bottom panel of
Fig.~\ref{fig.REZ}). Clearly, the 9:2 resonance cannot play any role
in these models.

Similar analysis was done for all other possible half-integer
resonances which centres fall within the computed model sequences
(e.g. for the other three resonances which loci are plotted in
Fig.~\ref{fig.diags}), as well as for all model sequences adopting
the convective parameters of sets \setj\ and \setzeo. Only for the
3:2 resonance the mismatch parameter is very small
($|\Delta_{3:2}|<0.05$) for all the models that show the period
doubling. It clearly indicates that the 3:2 resonance plays a key
role in causing the phenomenon. The question whether other
resonances can play an additional role for some particular models, for which
respective mismatch parameter is also close to $0$, cannot be
answered beyond doubt. It seems however, that this is not the case.
The period doubling domains as depicted in Fig.~\ref{fig.REZ}
through the amplitude of the sub-harmonic frequency, $B_{1/2}$, seem
to be void of features correlated with the location of the centres
of the other half-integer resonances (see also Fig.~\ref{fig.A}).
The crucial role of the 3:2 resonance in causing  the period
doubling behaviour in BL~Her type models is also supported by the
earlier analysis of radiative models by \cite{bm92}. Using the
relaxation technique \cite{bm92} proofed the key role of the 3:2
resonance.

In our opinion, the presented results leave no doubt that the 3:2 resonance with the first overtone is responsible for the period doubling in the hydrodynamic models and first of all in \BLGX.

\subsubsection{The best model for \BLGX}\label{sec.lcc}

\begin{figure*}
\centering
\resizebox{\hsize}{!}{\includegraphics{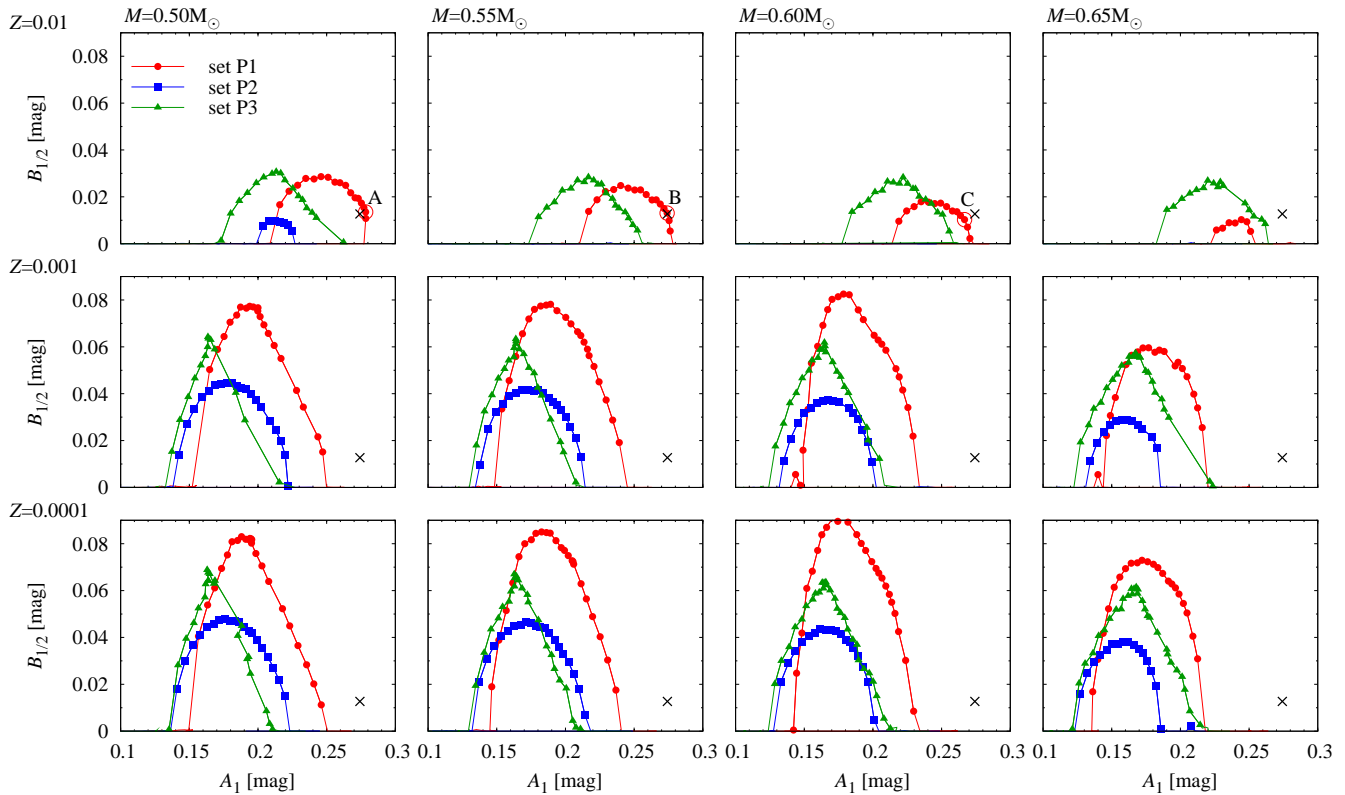}}
\caption{Amplitude of the sub-harmonic frequency component, $B_{1/2}$, plotted versus the amplitude, $A_1$, for the full grid of the computed non-linear models. Different sets of convective parameters (Table~\ref{tab.convpar}) were used in the computations, as indicated in the key in the top, left-most panel. Cross shows the amplitudes of \BLGX\ (see Section~\ref{sec.observations}). The best matching models of set \setp, A ($M=0.50\MS$, $Z=0.01$), B ($M=0.55\MS$, $Z=0.01$) and C ($M=0.60\MS$, $Z=0.01$), are encircled and labelled.}
\label{fig.A}
\end{figure*}

\begin{figure*}
\centering \resizebox{\hsize}{!}{\includegraphics{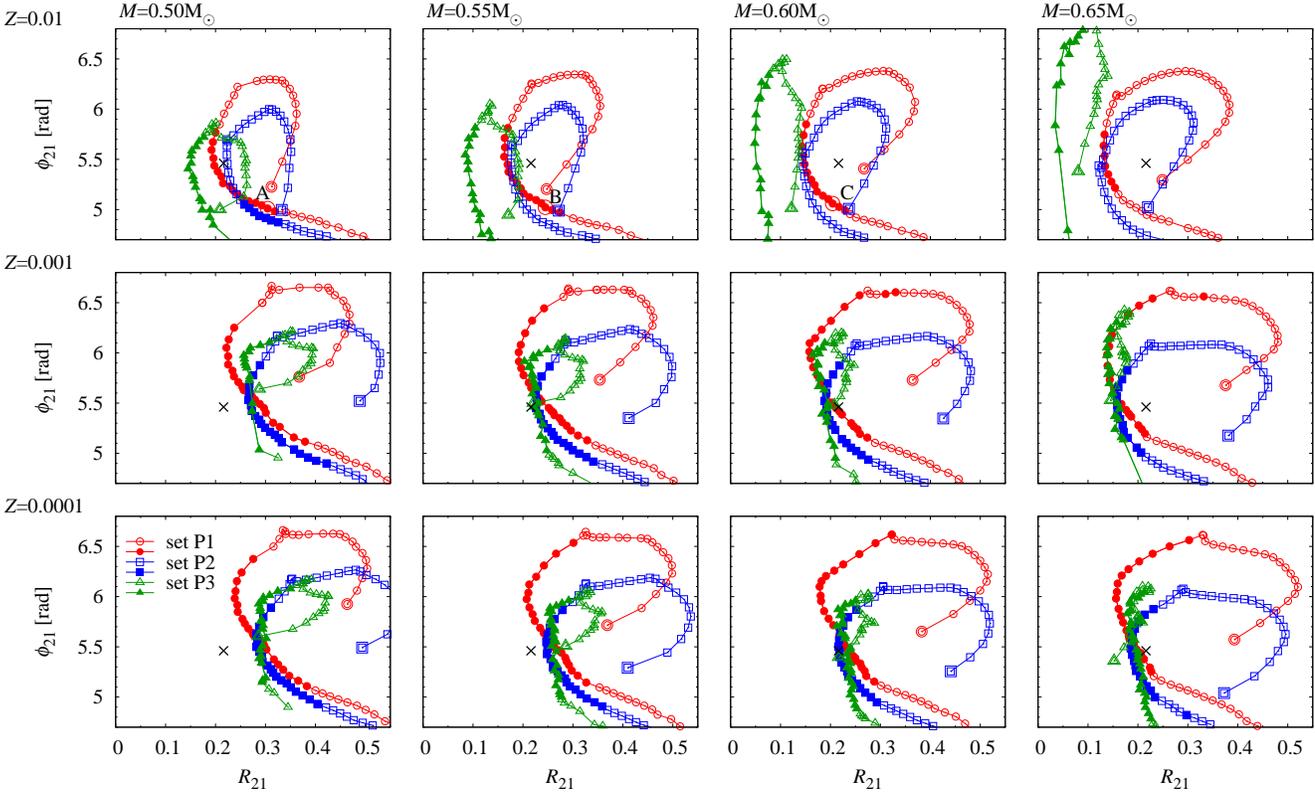}}
\caption{The phase difference, $\phi_{21}$, plotted versus the
amplitude ratio, $R_{21}$, for the full grid of the computed
non-linear models. Different sets of convective parameters
(Table~\ref{tab.convpar}) were used in the computations, as
indicated in the key in the bottom, left-most panel. Filled symbols
correspond to the models showing the period doubling. Crosses
represent the Fourier parameters of \BLGX\ (see
Table~\ref{tab.fourier}). The best matching models of set \setp, A,
B and C, are encircled and labelled.} \label{fig.RP}
\end{figure*}

In this Section we compare the computed model light curves with the light curve of \BLGX\ (Fig.~\ref{fig2}). Our comparison is based on the low-order Fourier decomposition parameters, the amplitudes, $A_1$ and $B_{1/2}$, the amplitude ratio, $R_{21}$, and the phase difference, $\phi_{21}$. Higher order parameters are modelled less accurately with the hydrocodes and are not used below. We note that physical parameters of all our models were adjusted to match the linear fundamental mode period of $2.4$\thinspace d with typical accuracy larger than $0.01$\thinspace d. Non-linear periods may differ only very slightly. These periods are close enough to the period of \BLGX\ and thus allow the direct comparison of the Fourier parameters which is done in Fig.~\ref{fig.A} ($A_1$ versus $B_{1/2}$) and in Fig.~\ref{fig.RP} ($\phi_{21}$ versus $R_{21}$). In these figures we show the results for all the computed models, for the full grid of model masses ($0.50\MS$, $0.55\MS$, $0.60\MS$ and $0.65\MS$, consecutive columns of Figs~\ref{fig.A} and \ref{fig.RP}) and model metallicities ($0.01$, $0.001$, and $0.0001$, consecutive rows of Figs~\ref{fig.A} and \ref{fig.RP}). Model sequences adopting different sets of convective parameters are plotted with different symbols, circles (set \setp), squares (set \setj) and triangles (set \setzeo). Filled symbols correspond to the models showing the period-doubling behaviour, while the models represented by open symbols show no sign of alternations. Only the former models are shown in Fig.~\ref{fig.A}. By default, the consecutive models in each model sequence differ by $2\LS$. Sometimes the difference is larger, as some models could not be converged to the limit cycle due to numerical problems (this is true particularly for set \setzeo). Cross in the figures shows the parameters of \BLGX\ (Tab.~\ref{tab.fourier}).

To find the best matching model for \BLGX\ we focus on the model
amplitudes (Fig.~\ref{fig.A}), the pulsation amplitude, $A_1$, and
the amplitude of period doubling, $B_{1/2}$. One could choose the
Fourier phases instead ($\phi_{21}$ and $\phi_{31}$), which little
depend on the pulsation amplitudes and are also not so sensitive to
the values of convective parameters \citep{rs09b}. We note however,
that one of the known problems in modelling the light curves with
pulsation hydrocodes is that computed Fourier phases are
systematically different from the observed phases. The problem
was studied in more detail in the RR Lyrae stars
\citep[e.g.][]{df99,kk98,f99}. It was also encountered in the
radiative and convective models of classical Cepheids
\citep[e.g.][]{mbm92,rs09b} and in radiative models of BL~Her stars
\citep{mb93a}. Origin of the discrepancy is not clear, although the
very simplified treatment of the radiation transfer in the
subphotospheric layers of all current hydrocodes is the most likely
culprit \citep{df99}.

Consequently
we choose to focus on the amplitudes first, even though they depend
on the convective parameters of the models stronger, and in the next
step, we compare the Fourier phases and amplitude ratios.

In Fig.~\ref{fig.A} we observe that in almost all model sequences
the period doubling domain is present. Only for three model
sequences, all with the largest considered metallicity ($Z=0.01$,
and $M=0.55-0.65\MS$), and adopting convective parameters of set
\setj, period doubling is not present. The other striking feature
visible in Fig.~\ref{fig.A} are large amplitudes of the period
doubling as compared to \BLGX\ and, at the same time, significantly
lower pulsation amplitudes, $A_1$. The systematics is clear. Lower
the mass and higher the metallicity of the models, higher the
pulsation amplitude and lower the amplitude of period doubling. Only
for the largest metallicity ($Z=0.01$) we can match the pulsation
amplitude of \BLGX\ (for sets \setp\ and \setzeo). Also for the most
metal rich model sequences the amplitude of period doubling is
comparable to what we observe in \BLGX. For lower metallicity models
($Z=0.001$ and $Z=0.0001$) the period doubling amplitude is much
higher through the whole period doubling domain and matches the
amplitude of \BLGX\ only at the domain edges. The large amplitude of
period doubling in our models and rather wide range of luminosities
over which period doubling domains extend (typically $30-40\LS$) is
somewhat surprising. This suggests that the period doubling should
be a rather common feature of BL~Her stars with periods around
$2.4$\thinspace d, which is not the case. We will address this
problem later in Section~\ref{sec.discussion}.

The above analysis of model amplitudes points that \BLGX\ should be of high metallicity. Then, we can simultaneously match its pulsation amplitude and the amplitude of period doubling. This is a likely possibility, as many BL~Her stars have high metallicities (see discussion in Section~\ref{sec.discussion}). However, one have to remember that model amplitudes depend on the convective parameters adopted in the computations, in particular on the mixing length and on the eddy-viscosity parameter. Comparison of the model results for sets \setp\ and \setj\ reveals the effect of changing the mixing length parameter (while keeping all other convective parameters fixed). Mixing length is equal to $1.5$ in our basic set \setp\ and is larger in set \setj\ ($\alpha=1.8$). The effect is clear, by increasing the mixing length the pulsation amplitude decreases as well as the amplitude of period doubling (which vanish for  $Z=0.01$ and three largest considered masses, Fig.~\ref{fig.A}). It seems that one can match the pulsation amplitude of \BLGX\ also for lower metallicities by the decrease of the mixing length below the value adopted in set \setp\ ($\alpha=1.5$). However at the same time the amplitude of period doubling, which is already very large, would increase even more. Also values of the mixing length below $1.5$ seem too small, typical values used in model computations being around $1.5-1.8$. Recent numerical simulations point to even larger values \citep{trampe}.

Mixing length parameter is usually fixed in model computations and it is the eddy viscosity parameter, which is commonly used to adjust the pulsation amplitudes in the convective models. It regulates the amount of viscous dissipation in the model and is a physically motivated counterpart of the artificial viscosity in the radiative models. Pulsation amplitudes of RR~Lyrae and Cepheid models clearly depend on that parameter \citep[see e.g.][]{rs09b}. Lower the eddy-viscous dissipation larger the pulsation amplitude. Also the existence and the amplitude of the period doubling computed in the convective RR~Lyrae models by \cite{kms11} depend on the eddy viscosity parameter.  Again, lower the eddy-viscous dissipation  larger the amplitude of the period doubling in case of the RR~Lyrae models. To match the pulsation amplitude of \BLGX\ for metallicities lower than $Z=0.01$ one has to decrease the eddy viscosity, which will also increase the amplitude of period doubling.

\begin{figure}
\centering
\resizebox{\hsize}{!}{\includegraphics{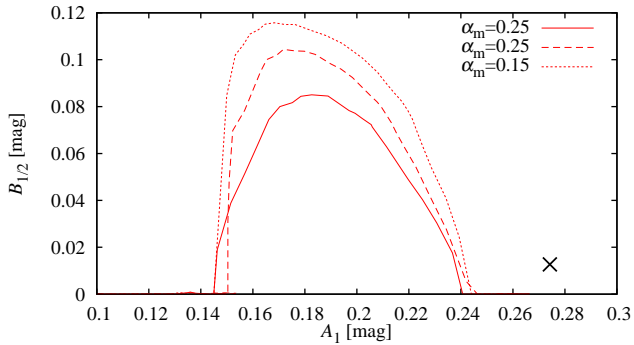}}
\caption{Amplitude of the sub-harmonic frequency component, $B_{1/2}$, plotted versus the amplitude, $A_1$, for the three model sequences, all with $M=0.55\MS$ and $Z=0.0001$,  adopting different values of the eddy viscosity parameter: $\alpha_{\rm m}=0.25$ (solid line), $\alpha_{\rm m}=0.20$ (dashed line) and $\alpha_{\rm m}=0.15$ (dotted line). All other parameters of the turbulent convection model are the same as in set \setp.}
\label{fig.ev}
\end{figure}

To test the effect of decreasing the eddy-viscosity parameter we have computed two additional model sequences, both with $M=0.55\MS$ and $Z=0.0001$, and with eddy-viscosity parameter equal to $\alpha_{\rm m}=0.20$ and $\alpha_{\rm m}=0.15$. The results are shown in Fig.~\ref{fig.ev} in which $B_{1/2}$ is plotted against $A_1$ for models of set \setp\ (solid line, $\alpha_{\rm m}=0.25$) and for two additional model sequences, with eddy-viscosity parameter decreased to $0.20$ (dashed lines) and $0.15$ (dotted line). We note that further decrease of eddy viscosity is not easy. Already with $\alpha_{\rm m}=0.20$  pulsation become very violent and we start to face the convergence difficulties. It is clear that amplitude of alternations strongly increase with the decreasing eddy viscosity parameter. However pulsation amplitude increased only very weakly -- much too small to match the amplitude of \BLGX.

Changes of the other convective parameters also affect the pulsation amplitudes which is usually compensated by the change of the eddy viscosity parameter. This is what we have done in set \setzeo, in which we consider the radiative cooling of the convective elements. Inclusion of radiative cooling makes the convection less efficient. Consequently the pulsation amplitudes increase which has to be compensated with the larger value of the eddy viscosity parameter. For the model sequences adopting convective parameters of set \setzeo\ we get the period doubling amplitudes higher than for set \setj\ and similar to that of set \setp. The pulsation amplitudes, $A_1$, are similar as in set \setj. Set \setzeo\ is interesting because it leads to significantly different light curve shapes than sets \setp\ and \setj, which we discuss in the next paragraphs.

Based on the model amplitudes we select three models, which match
the amplitudes of \BLGX, for the further, more detailed comparison.
The models, A, B and C, are encircled and labelled in
Fig.~\ref{fig.A}. All models adopt convective parameters of set
\setp, have $Z=0.01$ and differ in mas ($0.50\MS$ for model A,
$0.55\MS$ for model B and $0.60\MS$ for model C). Physical
parameters of these models are collected in the upper section of
Table~\ref{tab.models}. Before discussing these models in more
detail, we turn into the analysis of the low order Fourier
decomposition parameters, the $R_{21}$ amplitude ratio and the
$\phi_{21}$ phase difference. In Fig.~\ref{fig.RP} the two
quantities are plotted for the full grid of the computed models. The
range of variation of $R_{21}$ and $\phi_{21}$ is very large for the
considered models all of which have the same pulsation period
($\approx 2.4$\thinspace d). Relation between the two Fourier
parameters is very similar for sets \setp\ and \setj, which differ
in the value of the mixing length only, and is significantly
different for set \setzeo\ in which the effects of radiative losses
are turned on. The period doubling domains (filled symbols in
Fig.~\ref{fig.RP}) extend over a large range of Fourier phase,
$\phi_{21}$. There are period doubling models with Fourier phase
below $5.0$, as well as with Fourier phase well above $6.5$.
Consequently in all our model sequences we can point the model which
shows the period doubling and for which $\phi_{21}$ matches the
value of \BLGX. For the amplitude ratio we observe that period
doubling domains extend around the minimum of $R_{21}$ in the given
model sequence.   The simultaneous match of the \BLGX\ values of
$R_{21}$ and $\phi_{21}$ (cross in Fig.~\ref{fig.RP}) is possible
for several model sequences, which are located along the diagonal in
the mass-metallicity plane displayed in Fig.~\ref{fig.RP}. The best
match for the lowest metallicity sequences ($Z=0.0001$) is possible
for the highest masses ($0.60\MS$ and $0.65\MS$). For intermediate
metallicity ($Z=0.001$) we get the best match for the intermediate
masses ($0.55\MS$ and $0.60\MS$), and for the highest metallicity
($Z=0.01$) we get the best match for the lowest masses ($0.50\MS$).

Although for several models we can get an almost exact match
between the observed and model values of both $R_{21}$ and
$\phi_{21}$ parameters, only for the highest metallicity models we
can match the amplitudes of \BLGX. For lower metallicities
($Z=0.001$ and $Z=0.0001$) the pulsation amplitudes are lower than
observed and the amplitudes of alternations are much higher (see
Fig.~\ref{fig.A}).

The three models of $Z=0.01$, A, B and C, which match the amplitudes
of \BLGX\ best, are also labelled in Fig.~\ref{fig.RP}. These
models are relatively distant from the cross showing the location of
\BLGX\ and for the lower luminosity models from the same model
sequence the $R_{21}$ and $\phi_{21}$ can be matched much better.
Still the agreement for models A, B and C is reasonable, as
quantitative comparison presented in the lower section of
Table~\ref{tab.models} shows. For model C the $R_{21}$ amplitude
ratio is lower than observed by only $5$\thinspace per cent and the
$\phi_{21}$ phase difference is lower by $0.41$\thinspace rad. The
agreement is similar for model B and slightly worse for model A for
which the amplitude ratio is significantly different than observed.
In Table~\ref{tab.models} we also provide the comparison for higher
order Fourier parameters, $R_{31}$ and $\phi_{31}$. The differences
are very large. Model values of $R_{31}$ are much larger than
observed, even for the best model C, by as much as $85$\thinspace
per cent. For $\phi_{31}$ the lack of agreement is also remarkable,
the model phases being much lower than observed. Detailed analysis
of all our models points that although we can easily find models for
which two Fourier parameters (e.g. both phases) simultaneously match
the observed values, the simultaneous match of all low order Fourier
parameters (amplitudes, amplitude ratios and phases) is not
possible.

\begin{table}
\caption{Parameters of the best matching models for \BLGX. In the top section of the Table physical parameters are provided and in the bottom section of the Table the Fourier decomposition parameters are given. The differences with respect to \BLGX\ (model$-$star) are given in the parentheses (relative differences for amplitudes and amplitude ratios).}
\label{tab.models}
\centering
\setlength{\tabcolsep}{5pt}
\begin{tabular}{lccc}
\hline
model: & A & B & C\\
\hline
\multicolumn{4}{c}{{\it physical parameters}}\\
\hline
$M/\MS$ & 0.50 & 0.55  & 0.60 \\
$L/\LS$ & 168.0& 176.0 & 184.0\\
$T_{\rm eff}$ [K] & 6241.6 & 6201.4 & 6167.2 \\
\hline
\multicolumn{4}{c}{{\it Fourier parameters}}\\
\hline
$A_1$       & $0.2784\,(+1.5\%)$ &  $0.2740\,(-0.1\%)$ & $0.2667\,(-2.7\%)$ \\
$B_{1/2}$   & $0.0135\,(+6.5\%)$ &  $0.0131\,(+2.8\%)$ & $0.0103\,(-19.1\%)$ \\
$R_{21}$    & $0.304\,(+40.7\%)$ &  $0.245\,(+13.4\%)$ & $0.205\,(-5.1\%)$ \\
$\phi_{21}$ & $5.010\,(-0.451)$  &  $5.022\,(-0.439)$  & $5.060\,(-0.401)$ \\
$R_{31}$    & $0.236\,(+211\%)$  &  $0.185\,(+143\%)$  & $0.140\,(+84\%)$ \\
$\phi_{31}$ & $3.376\,(-0.618)$  &  $3.054\,(-0.940)$  & $2.709\,(-1.285)$ \\
\hline
\end{tabular}
\end{table}

\begin{figure}
\centering
\resizebox{\hsize}{!}{\includegraphics{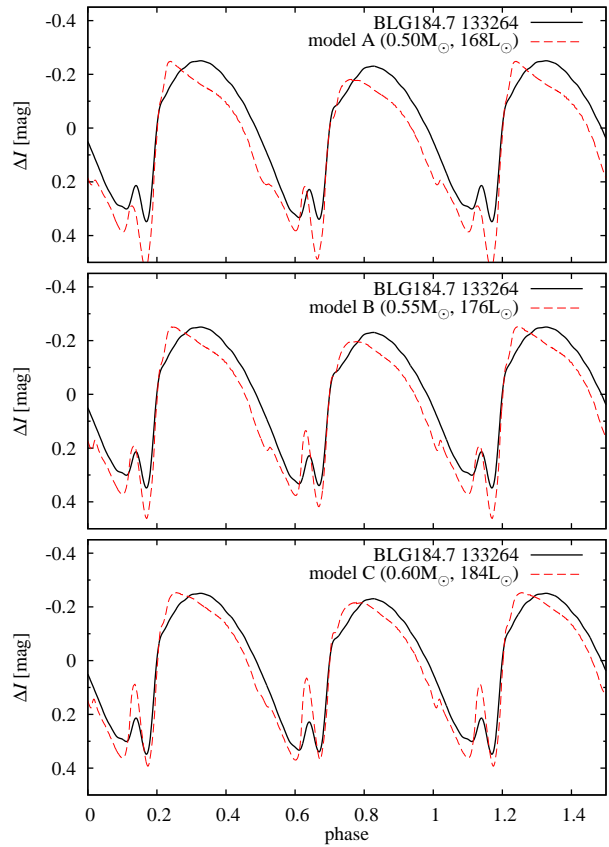}}
\caption{Comparison of the observed $I$-band light curve of \BLGX\ (solid line) with the best matching models of $0.50\MS$ (top panel), $0.55\MS$ (middle panel) and $0.60\MS$ (bottom panel). The model light curves were shifted vertically to match the maximum brightness of \BLGX\ and shifted in phase so the ascending branch of the models and observed light curve overlap.}
\label{fig.lcc}
\end{figure}

In Fig.~\ref{fig.lcc} we directly compare the $I$-band light curves
of \BLGX\ (Fig.~\ref{fig2}) and the three selected models A, B and
C. The model light curves were shifted vertically in order to match
the maximum brightness of \BLGX, and shifted in phase, so the
ascending branches of the light curves overlap. The overall light
curve shape of \BLGX\ is nicely reproduced by all three models, but
several discrepancies are apparent. In the observed light curve,
after the pronounced shoulder on the ascending branch the light
raise becomes less steep. The observed light curve is rounded at the
maximum light. The shoulder is also visible in the models however,
the ascending branch is steep all the time till the maximum
brightness which consequently occurs at earlier phases in the
models. The light fall also begins earlier and as a result the
overall width of the light curve is slightly smaller in the models.
At brightness minimum the pronounced bump is clearly visible in both
the observations and in the models. Its location is well reproduced
by the models, but its amplitude in the models is much higher. The
changing character of the brightness minima around the bump during
the alternations (each second minimum after the bump is deeper) is
qualitatively reproduced by the models. Clearly we get the best
agreement for model C, which is also confirmed by more quantitative
comparison (Table~\ref{tab.models}), but the agreement for the other
two models is reasonable, as well.

The discrepancies between the model and the observed light
curves are not very serious. They concern only the more subtle
features of the light curve, like the detailed shape of the
brightness maximum or minimum. Such discrepancies are reflected
predominantly in the higher order Fourier parameters. Accurate
modeling of these higher order Fourier terms is a very challenging
problem, though. Of particular importance is the treatment of
radiative transfer in the outer zones of the model and the
transformation of the bolometric light curve to the $I$-band light
curve. These are both treated in a very approximate way in our code
(see Section~\ref{sec.numericalmethods}). Therefore, we do not
expect that the models will satisfactorily reproduce all details of
the observed light curves. This is a well known deficiency, common
to all hydrodynamic models of large amplitude radial pulsators
\citep[e.g.,][]{kk98}.

Comparison of the observed and model light curves for \BLGX\ point that the star should be of relatively high metallicity ($Z\approx 0.01$) and its mass should be lower than $0.60\MS$. Its luminosity should be large around $170-180\LS$. In the next Section we confront these parameters with the results of recent horizontal branch evolutionary models.

\section{Comparison with the evolutionary tracks}\label{sec.evol}

The evolutionary status of BL~Her stars was discussed in a series of papers by Gingold \citep[see][for a summary]{g85}. Since then, several groups published the horizontal branch (HB) evolutionary tracks for extensive grid of models covering a large range of masses and chemical compositions. These include e.g. ${\rm Y}^2$ tracks by \cite{Y2}, Padova tracks by \cite{padova}, the BaSTI tracks by \cite{BaSTI06} and the Dartmouth tracks by \cite{dartmouth}. Below we briefly review the Gingold's results and compare them with the outcome of more recent evolutionary computations. To this aim we use the BaSTI and Dartmouth tracks. Based on the location of the BL~Her instability strip in the HR diagram we conclude on the possible masses of BL~Her stars as predicted by the evolution theory. Finally, a comparison with the parameters of our best fitting models for \BLGX\ (Section~\ref{sec.lcc}) is done.

\begin{figure*}
\centering \resizebox{\hsize}{!}{\includegraphics{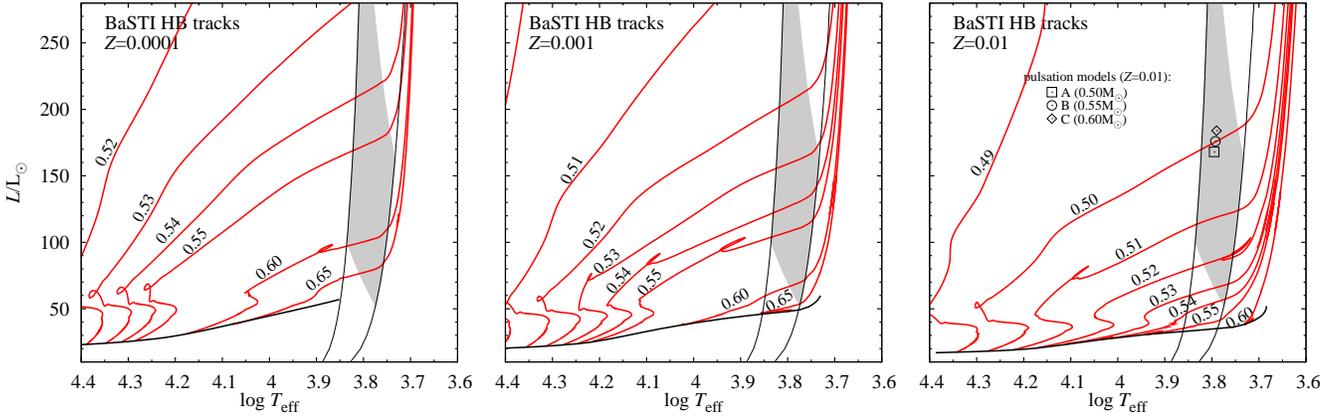}}
\caption{The BaSTI \citep{BaSTI06} horizontal branch evolutionary
tracks for three values of model's metallicity, $Z=0.0001$ in the
left-hand panel, $Z=0.001$ in the middle panel and $Z=0.01$ in the
right-hand panel. Evolutionary tracks start at the ZAHB, which is
marked with the solid, horizontally running line. Each track is
labelled with the corresponding model mass. Over-plotted are the
edges of the instability strip, with the shaded area indicating the
BL~Her domain, with fundamental mode periods between 1\thinspace d
(lower boundary of the shaded area) and 4\thinspace d (upper
boundary of the shaded area).} \label{fig.evBa}
\end{figure*}

\begin{figure*}
\centering \resizebox{\hsize}{!}{\includegraphics{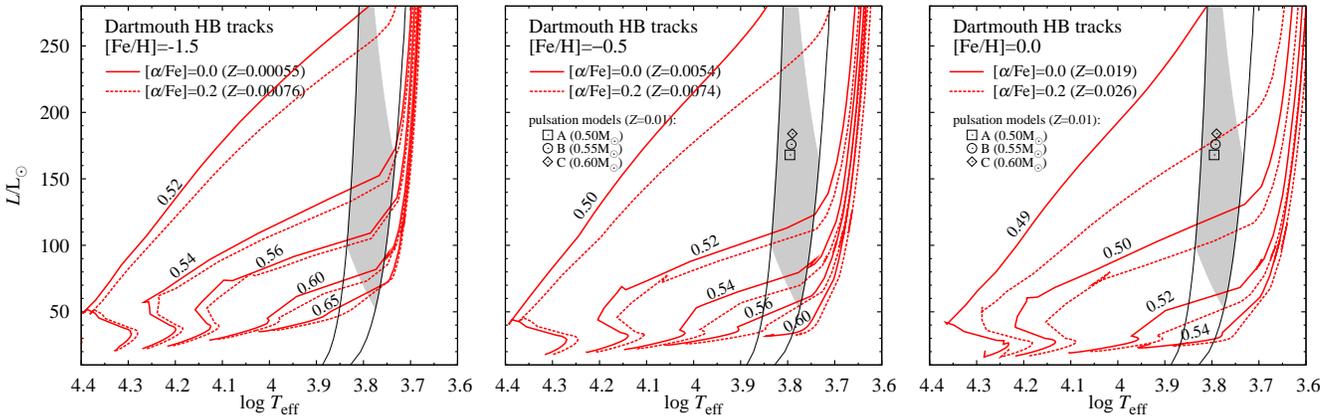}}
\caption{The Dartmouth \citep{dartmouth} horizontal branch
evolutionary tracks for three values of model's iron abundance,
${\rm [Fe/H]}=-1.5$ in the left-hand panel, ${\rm [Fe/H]}=-0.5$ in
the middle panel and ${\rm [Fe/H]}=0.0$ in the right-hand panel.
The evolutionary tracks are labelled with the corresponding model
mass. In each panel we also plotted additional set of evolutionary
tracks for models assuming enhanced abundance of the
$\alpha$-elements, $[\alpha/{\rm Fe}]=+0.2$ (dashed lines).
Over-plotted are the edges of the instability strip, with the shaded
area indicating the BL~Her domain, with fundamental mode periods
between 1\thinspace d (lower boundary of the shaded area) and
4\thinspace d (upper boundary of the shaded area).} \label{fig.evDa}
\end{figure*}

Modelling of the post-main sequence evolution of the low-mass stars
is particularly difficult \citep[see e.g.][]{chiosi92}. During the
evolution along the red giant branch (RGB) these stars develop a
degenerate helium core. Then mass of the core reaches a critical
value the runaway helium burning (so called helium flash) is
initiated. This violent and rapid phase of the evolution,
accompanied by a substantial mass-loss, cannot be followed with the
current evolutionary codes. Instead the evolution is restarted at
the Zero-Age Horizontal Branch (ZAHB), at much larger effective
temperatures. Progenitors of the BL~Her stars start their evolution
at the blue side of the ZAHB, far beyond the blue edge of the
instability strip. During the core helium burning, which is a
relatively long phase of the evolution, they remain close to the
ZAHB. As the helium becomes gradually depleted in the core, the
evolution becomes faster and star moves towards the red, crossing
the instability strip and thus, becoming a BL~Her variable.
According to Gingold evolutionary computations two scenarios are
possible, which depend mostly on the mass of the envelope \citep[see
fig.~1 in][]{g85}. In the first scenario, the star crosses the
instability strip once and evolves away up along the asymptotic
giant branch (AGB). This is the scenario which one can also see in
Figs~\ref{fig.evBa} and \ref{fig.evDa}. In the second scenario, so
called `bluenose' is possible. Due to the structure changes between
the helium and hydrogen burning shells, after approaching the AGB
the evolution is directed to the blue and then, back towards the
AGB. Two additional crossings of the instability strip are then
possible before the star finally climbs-up along the AGB. In both
scenarios the star can enter the instability strip again, during the
post-AGB evolution (while executing the thermal pulses), becoming a
W~Vir or RV~Tau-type variable.

Both types of tracks can be found in the more recent evolutionary computations, however, the `bluenose' behaviour is very rare. It occurs in a narrow range of masses, and only for larger than solar metallicities (BaSTI, Dartmouth) and in addition in models assuming a large enhancement of the $\alpha$-elements ($[{\rm \alpha/Fe]}>+0.6$, Dartmouth). We do not discuss this scenario in the following.

The recent evolutionary tracks are displayed in Figs~\ref{fig.evBa}
(BaSTI) and \ref{fig.evDa} (Dartmouth). In these figures the edges
of the instability strip are marked with solid, vertically running
lines. The grey-shaded area in between corresponds to the BL~Her
pulsation domain with fundamental mode periods between 1\thinspace d
(lower boundary of the shaded area) and 4\thinspace d (upper
boundary of the shaded area)\footnote{Location of the instability
strip is approximate. It was computed assuming $M=0.55\MS$,
$Z=0.001$ and convective parameters of set~\setp.}. Below that
region the RR~Lyrae instability strip extends. The following
discussion is based on the BaSTI tracks. In the three panels of
Fig.~\ref{fig.evBa} we have plotted the evolutionary tacks for three
different metallicities, $Z=0.0001$ in the left-hand panel,
$Z=0.001$ in the middle panel and $Z=0.01$ in the right-hand panel.
Several evolutionary tracks for different model masses are plotted.
It is clearly visible that for a given metallicity, the masses of
models which can enter the BL~Her domain during the evolution are
restricted. First, consider the lowest metallicity models
($Z=0.0001$, left-hand panel of Fig.~\ref{fig.evBa}). Model with the
largest mass displayed in the figure ($M=0.65\MS$) cross the BL~Her
domain at its lower part at $L<80\LS$. Lower the model mass, higher
the luminosity at the crossing with the BL~Her domain. The models
with $M\ge 0.54\MS$ enter the instability strip. For lower masses,
crossing with the instability strip occurs at larger luminosities,
above the BL~Her domain, or does not occur at all (evolution along
almost vertical tracks or even to the blue). As the metallicity is
increased evolutionary tracks for models of a given mass (their
pre-AGB part) shift towards the red and bend towards the lower
luminosities. Consequently the less massive models can enter the
BL~Her domain from the top. On the other hand, the tracks for most
massive models leave the BL~Her domain running below it. For
$Z=0.001$ (middle panel of Fig.~\ref{fig.evBa}) the tracks for
models with $0.52\MS\le M\le 0.60\MS$ cross the instability strip
and for $Z=0.01$ (right-hand panel of Fig.~\ref{fig.evBa}) the range
of possible masses is even smaller, $0.50\MS\le M\le 0.54\MS$. All
tracks with $M>0.54\MS$ run below the BL~Her domain or even start
beyond the red edge of the instability strip ($0.60\MS$).

 Qualitatively the same scenario is evident from the analysis of the
Dartmouth evolutionary tracks plotted in Fig.~\ref{fig.evDa}. The
one-to-one comparison with the BaSTI tracks is difficult, because
the Dartmouth tracks were computed for specific values of [Fe/H],
which do not correspond to the $Z$ values adopted in the BaSTI
tracks. In Fig.~\ref{fig.evDa} we have also plotted additional set
of evolutionary tracks for models adopting slightly enhanced
abundance of the $\alpha$-elements. Detailed list of the elements
with their abundances can be found in table~2 of \cite{dartmouth}.
Their table~3 provides the corresponding metallicity values, $Z$,
which we also provide in Fig.~\ref{fig.evDa}. Just as for the BaSTI
tracks we observe that lower the metallicity, larger the possible
masses (and larger their range) of models crossing the BL~Her domain
during the evolution. The main effect of increasing the abundance of
the $\alpha$-elements is shift of the tracks towards the lower
luminosities and slightly lower effective temperatures.
Consequently, for the models with increased abundance of the
$\alpha$ elements the range of possible masses that allow for
crossing the BL~Her domain (at given $Z$) shifts towards slightly
smaller mass values.

Presented results are also in qualitative agreement with the ${\rm Y}^2$ \citep{Y2} evolutionary models as well as with the Padova \citep{padova} HB tracks (for $M\ge 0.55\MS$; tracks assuming lower $M$ are not present in the Padova database).

As discussed above, the evolution theory put constraints on the
possible masses of the BL~Her stars, which depend on the
metallicity. Larger the metallicity, lower the possible masses of
the BL Her stars. This has severe consequences for our best matching
models for \BLGX\ discussed in Section~\ref{sec.lcc}, A, B and C,
all of which have $Z=0.01$. The models are shown with different
symbols in Figs~\ref{fig.evBa} and \ref{fig.evDa}. Comparison is
straightforward for the BaSTI tracks, as these were computed also
for $Z=0.01$, and is done in the right-hand panel of
Fig.~\ref{fig.evBa}. We clearly see that models B ($M=0.55\MS$) and
C ($M=0.60\MS$) cannot be matched by the respective evolutionary
tracks. Tracks for these masses run below the BL~Her domain. To the
contrary, model A is in excellent agreement with the evolutionary
computations, the track for model of $0.50\MS$ nearly crosses the
square representing model A in the Figure. The one-to-one comparison
with the Dartmouth tracks is somewhat difficult as metallicity of
our best models, $Z=0.01$, falls in between ${\rm [Fe/H]}=0.0$ and
${\rm [Fe/H]}=-0.5$ for which the Dartmouth tracks were computed.
Consequently, we have plotted the models in both the middle
($Z\approx 0.0054$) and right-hand panel ($Z\approx 0.019$) of
Fig.~\ref{fig.evDa}. Metallicities are somewhat larger for
evolutionary models assuming increased abundance of the
$\alpha$-elements (see labels in Fig.~\ref{fig.evDa}). Conclusion is
the same as for the BaSTI tracks. Only the least massive model A
($M=0.50\MS$) can be matched with the evolutionary tracks.

Luminosities of models B and C are much too high for their masses if confronted with the evolutionary tracks. Consequently, model A is our best model for \BLGX, selected based on both the pulsation computations and the evolution theory.

\section{Discussion and Conclusions}\label{sec.discussion}

In this paper we report the discovery of the first BL~Her star
showing the regular alternations of the light curve -- the period
doubling. The star was discovered nearly twenty years after
existence of such objects was predicted by \cite{bm92}, who found
the period doubling behaviour in their radiative, BL~Her-type
models. The star, \BLGX, was found in the Galactic bulge OGLE-III
survey data. It pulsates in the fundamental mode, with period equal
to 2.4\thinspace d. Its Fourier decomposition parameters firmly
place it among BL~Her variables. We also identified another
strong period doubling candidate (\BLGY) in which however, the
amplitude of alternations is very low and more accurate observations
are necessary to confirm the detection.

Discovery of period doubling in a BL~Her star provides a
motivation for in-depth theoretical study of the phenomenon. In
particular, the computation of new models, with up to date physics
and including the convective energy transfer is of great value. In
this paper we present the results of initial model survey, which is
focused on reproducing the period doubling behaviour in \BLGX. Thus,
all computed models have $P_0\approx 2.4$\thinspace d and lie along
an approximately straight line in the HR diagram. We have found a
large domain, extending over a range of few tenths of solar
luminosities, in which period doubling behaviour is possible, with
rather large amplitude of alternations, of order of several hundreds
of magnitude. The period ratios of the models showing the period
doubling effect, indicate the crucial role of the 3:2 resonance
between the fundamental mode and the first overtone in causing the
period doubling behaviour, confirming the earlier result of
\cite{bm92}.

The model light curves were transformed into the $I$-band and
compared with the observed light curve of \BLGX. This was done
with the intermediary of the Fourier decomposition parameters. We
were able to match the observed pulsation amplitude and the
amplitude of the period doubling only for relatively high
metallicity, $Z=0.01$. For three considered model masses, $0.50\MS$,
$0.55\MS$ and $0.60\MS$, the model light curves reproduce the
observed light curve quite well (Fig.~\ref{fig.lcc}). Low order
Fourier parameters, $R_{21}$ and $\phi_{21}$, are reproduced
satisfactorily (particularly for $M=0.60\MS$). Also the secondary
features of the light curve, in particular the pronounced bump at
the minimum light are qualitatively reproduced. However, only the
lowest mass model ($M=0.50\MS$) agree with the stellar evolution
theory. For $Z=0.01$ evolutionary tracks for models of $M\ge
0.60\MS$ run well beyond the red edge of the BL~Her domain.
Consequently, our best estimates for the parameters of \BLGX\ are
those of model A (Table~\ref{tab.models}), specifically
$Z=0.01$, $M=0.50\MS$ and $L=168\LS$.

How robust are these estimates? Uncertainties lie in both the
pulsation theory and in the evolution theory. Our conclusion about
the high metallicity of \BLGX\ is based on the pulsation amplitudes
of the computed models. These depend on the adopted convective
parameters, in particular on the mixing-length parameter, $\alpha$,
and the eddy viscosity parameter, $\alpha_{\rm m}$. As we discussed
in Section~\ref{sec.lcc}, change of these parameters is not likely
to affect our results strongly. Nevertheless, more extensive model
survey, including the variation of other parameters of the turbulent
convection model, is needed and planned. In particular, the
effect of including the turbulent pressure should be examined.

Derived metallicity can, in principle, be larger than $Z=0.01$, but
only slightly. In test computations with $Z=0.02$ and
assuming convective parameters of set \setp, we find no traces of
period doubling behaviour (independently of the model mass).

At first glance, metallicity of order of $Z=0.01$ may seem too
high for the Population~II object. The metal rich stars, however,
are not exceptional in the Galactic bulge. Photometric
metallicity estimates for several Type II Cepheids in the bulge
range from $-1.4$ to $+0.6$, with mean value of ${\rm [Fe/H]} =
-0.6\pm 0.17$ \citep{hw84, wallerstein}. Spectroscopic metallicity
determinations, available mostly for bright K and M giants and
Red Clump stars, indicate a large spread of metallicities in the
bulge, too, with ${\rm [Fe/H]}$ varying from $-1.3$ to $+0.5$
\citep[see][for recent review]{zoccali}. ${\rm [Fe/H]}$
determination for several main sequence stars in the bulge was also
possible, during the microlensing events \citep{bensby}. For these
stars, metallicity range form $-0.7$ to $+0.5$. Metallicity can also
be estimated for the RR Lyrae stars pulsating simultaneously in the
fundamental and the first overtone modes (RRd stars). The OIII-CVS
\citep{sosz11} lists several bulge RRd stars with unusually short
periods and low period ratios, down to $P_0\approx 0.35$\thinspace d
and $P_1/P_0\approx 0.726$. The existence of these stars may be
explained by high metal abundances, up to ${\rm [Fe/H]}\approx
-0.36$. As $Z=0.01$ corresponds to ${\rm [Fe/H]}\approx-0.27$, we
see that our estimate of metallicity of \BLGX\ is not unreasonable.

For the luminosity, we can firmly indicate its lower limit. The
period doubling domain in the HR diagram always coincide with the
crossing of the 3:2 resonant line with the $P_0=2.4$\thinspace d
line. Run of the two lines shows very little sensitivity to
the values of the convective parameters (Section~\ref{sec.linear}),
as well as to the mass and metallicity of the models
(Fig.~\ref{fig.diag5}). The lowest luminosity at which the crossing
occurs is $\approx 150\LS$.

Based on the recent evolutionary tracks, we have constrained the
mass of \BLGX. With metallicity of $Z=0.01$ only the models in a
narrow mass range of $\approx 0.50-0.53\MS$ evolve through the
BL~Her pulsation domain. We get the best match with the
pulsation models for $M=0.50\MS$. We note however, that at $Z=0.01$
the evolutionary tracks that cross the BL~Her domain vary rapidly
with mass of the models. Although the current evolutionary
tracks of several groups lead to the same conclusions
(Section~\ref{sec.evol}), the systematic errors in the micro-physics
(e.g. opacities) or in the treatment of physical phenomena (e.g.
convection, rotation, mass loss) may effect the results of
evolutionary computations and somewhat change the best matching
mass for \BLGX.

Hydrodynamic model surveys \citep[][this work]{bm92} indicate that
period doubling phenomenon should be a rather common property of the
BL~Her stars, at least of those with periods close to
$2.4$\thinspace d. \cite{bm92} found the period doubling behaviour
in a period range of $2.0-2.6$\thinspace d. Predicted amplitudes of
alternations are usually large, much larger than
observed in \BLGX. Thus, alternations should be easily
detectable. Currently, around 250 BL~Her variables are known;
55 stars are listed in GCVS \citep{gcvs} and close to 200 in
OIII-CVS \citep[][2011, in prep.]{sosz08,sosz10}.  OGLE observations
are accurate enough to discover alternations with amplitudes as
low as a few mmag. Hence, it is somewhat surprising that \BLGX\ is,
at the moment, the only BL~Her star with firmly detected period
doubling behaviour. We note, that our strong period doubling
candidate, \BLGY, also has a period in the expected range. On the
other hand, only few of the known BL~Her stars have periods
close to 2.4\thinspace d. Our study shows that higher
metallicity (e.g. close to solar, which may not be exceptional for
BL~Her stars) may inhibit the period doubling effect.
Definitely, more detailed non-linear study is needed.

In the forthcoming paper we plan to study the full grid of models,
covering the whole BL~Her domain in the HR diagram, not only the
2.4\thinspace d models. Such survey will reveal the structure of the
period doubling domain in the HR diagram, in particular the periods
at which the phenomenon is possible and factors which control the
existence and strength of the period doubling effect.

\section*{Acknowledgments}

Model computations presented in this paper have been conducted on
the psk computer cluster in the Copernicus Centre, Warsaw, Poland.
RS is supported by the Austrian Science Fund (FWF project AP
21205-N16). The research leading to these results has also received
funding from the European Research Council under the European
Community's Seventh Framework Programme (FP7/2007-2013)/ERC grant
agreement no. 246678.


\label{lastpage}

\end{document}